\documentclass[longauth]{aa}  
\usepackage{graphicx}
\usepackage{txfonts}
\usepackage{ulem}
\usepackage{multirow}

\newcommand{\kps}[0]{\ensuremath{\mathrm{km\,s}^{\mathrm{-1}}}}
\newcommand{\grvs}[0]{\ensuremath{G_\mathrm{RVS}}}
\newcommand{\vsini}[0]{\ensuremath{V\sin{i}}}
\newcommand{\teff}[0]{\ensuremath{T_{\mathrm{eff}}}}
\newcommand{\logg}[0]{\ensuremath{\log g}}
\newcommand{\feh}[0]{\ensuremath{\mathrm{[Fe/H]}}}
\newcommand{\bprp}[0]{\ensuremath{G_{\mathrm{BP}}-G_{\mathrm{RP}}}}

\newcommand{\vmacro}[0]{\ensuremath{V_\mathrm{macro}}}

\newcommand{\elena}[1]{{#1}}
\newcommand{\corr}[1]{#1}
\newcommand{\Gaia}[0]{\textit{Gaia}}
\newcommand{\GDRthree}[0]{\Gaia\ DR3}

\providecommand{\orcit}[1]{\protect\href{https://orcid.org/#1}{\protect\includegraphics[width=8pt]{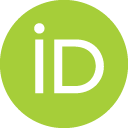}}}

\usepackage{xcolor}
\usepackage{hyperref}
\hypersetup{
    colorlinks = true,
    linkcolor = blue,
    anchorcolor = blue,
    citecolor = blue,
    filecolor = blue,
    urlcolor = blue
}
\def\DraftCommentColor{blue!80}
\providecommand{\draftcomment}[1]{\iftrue{\color{\DraftCommentColor}{#1}}\fi
}

\makeatletter
\DeclareRobustCommand*{\fieldName}[1]{%
    \begingroup\@fieldName\scantokens{\texttt{\small
    {#1}}\noexpand}\endgroup}
\begingroup\lccode`\~=`\_\relax
\lowercase{\endgroup\def\@fieldName{\catcode`\_=\active \let~\_}}
\makeatother

\providecommand{\linktodoc}{https://gea.esac.esa.int/archive/documentation/GDR3}  
\providecommand{\linktoparam}[2]{\href{\linktodoc/Gaia_archive/chap_datamodel/sec_dm_main_source_catalogue/ssec_dm_#1.html\##1-#2}{\fieldName{#2}\xspace}}

\providecommand{\linktoaps}[1]{\href{\linktodoc/Gaia_archive/chap_datamodel/sec_dm_astrophysical_parameter_tables/ssec_dm_astrophysical_parameters.html\#astrophysical_parameters-#1}{\fieldName{#1}\xspace}}

\newcommand{\vbroad}[0]{{\tt vbroad}}

\begin{document}

   \title{\Gaia\ Data Release 3: Properties of the line broadening parameter derived with the Radial Velocity Spectrometer (RVS)}

   \titlerunning{\GDRthree\: Properties of the line broadening parameter derived with the RVS}

   \author{Y. Fr\'emat\orcit{0000-0002-4645-6017}
          \inst{\ref{inst:01}}\thanks{\email{yves.fremat@observatory.be}}
          \and
          F. Royer\orcit{0000-0002-9374-8645}\inst{\ref{inst:02}}\thanks{\email{frederic.royer@obspm.fr}}
          \and
          O. Marchal\inst{\ref{inst:08}}
      \and
      R.~Blomme\inst{\ref{inst:01}}
   	  \and P.~Sartoretti\inst{\ref{inst:02}}
   	  \and A.~Guerrier\inst{\ref{inst:03}}
      \and P.~Panuzzo\inst{\ref{inst:02}}
      \and D.~Katz\inst{\ref{inst:02}}
   	  \and G. M.~Seabroke\inst{\ref{inst:04}}
      \and F.~Th\'{e}venin\inst{\ref{inst:05}}
      \and M.~Cropper\inst{\ref{inst:04}}
      \and K.~Benson\inst{\ref{inst:04}}
      \and Y.~Damerdji\inst{\ref{inst:06},\ref{inst:07}}
      \and  R.~Haigron\inst{\ref{inst:02}}
      \and A.~Lobel\inst{\ref{inst:01}}
      \and M.~Smith\inst{\ref{inst:04}}
      \and S.G.~Baker\inst{\ref{inst:04}}
      \and L.~Chemin\inst{\ref{inst:08ter}}
      \and M.~David\inst{\ref{inst:09}}
      \and C.~Dolding\inst{\ref{inst:04}}
      \and E.~Gosset\inst{\ref{inst:07},\ref{inst:10}}
      \and K.~Jan{\ss}en\inst{\ref{inst:11}}
      \and  G.~Jasniewicz\inst{\ref{inst:12}}
      \and G.~Plum\inst{\ref{inst:02}}
      \and N.~Samaras\inst{\ref{inst:01},\ref{inst:12bis}}
      \and O.~Snaith\inst{\ref{inst:02}}
      \and  C.~Soubiran\inst{\ref{inst:13}}
      \and O.~Vanel\inst{\ref{inst:02}}
\and J.~Zorec\inst{\ref{inst:17}}       
\and T.~Zwitter\inst{\ref{inst:14}}
      \and N.~Brouillet\inst{\ref{inst:13}}
      \and E.~Caffau\inst{\ref{inst:02}}
      \and F.~Crifo\inst{\ref{inst:02}} 
      \and C.~Fabre\inst{\ref{inst:03}}
      \and F.~Fragkoudi\inst{\ref{inst:0018},\ref{inst:0019}}
      \and H.E.~Huckle\inst{\ref{inst:04}}
      \and Y.~Lasne\inst{\ref{inst:03}}
      \and N.~Leclerc\inst{\ref{inst:02}}
      \and  A.~Mastrobuono-Battisti\inst{\ref{inst:02}}
\and A.~Jean-Antoine Piccolo\inst{\ref{inst:03}}      
      \and Y.~Viala\inst{\ref{inst:02}}
                }

   \institute{
   	Royal Observatory of Belgium, Avenue circulaire 3, B-1180 Bruxelles, Belgium
   	\label{inst:01}
   	\and  GEPI, Observatoire de Paris, Universit\'{e} PSL, CNRS, 5 Place Jules Janssen, F-92190 Meudon, France\relax
   	\label{inst:02}
   	\and Observatoire astronomique de Strasbourg, Universit\'{e} de Strasbourg, CNRS, 11 rue de l'Universit\'{e}, F-67000 Strasbourg, France
   	\label{inst:08}
   	\and CNES Centre Spatial de Toulouse, 18 avenue Edouard Belin, F-31401 Toulouse Cedex 9, France\relax                                                
   	\label{inst:03}
   	\and
   	Mullard Space Science Laboratory, University College London, Holmbury St Mary, Dorking, Surrey, RH5 6NT, United Kingdom\relax
   	\label{inst:04}
   	\and Universit\'{e} C\^{o}te d’Azur, Observatoire de la C\^{o}te d’Azur, CNRS, Laboratoire Lagrange, Boulevard de l’Observatoire, CS 34229, 06304 Nice, France
   	\label{inst:05}
   	\and CRAAG - Centre de Recherche en Astronomie, Astrophysique et G\'{e}ophysique, Route de l'Observatoire, Bp 63 Bouzareah, DZ-16340, Alger, Alg\'{e}rie\relax                                               
   	\label{inst:06}
   	\and Institut d'Astrophysique et de G\'{e}ophysique, Universit\'{e} de Li\`{e}ge, 19c, All\'{e}e du 6 Ao\^{u}t, B-4000 Li\`{e}ge, Belgium\relax
   	\label{inst:07}
   	 \and Centro de Astronom\'ia, Universidad de Antofagasta, Avda. U. de Antofagasta, 02800 Antofagasta, Chile
   	\label{inst:08ter}
   	\and Universiteit Antwerpen, Onderzoeksgroep Toegepaste Wiskunde, Middelheimlaan 1, B-2020 Antwerpen, Belgium\relax                                                                                         
   	\label{inst:09}
   	\and F.R.S.-FNRS, Rue d'Egmont 5, B-1000 Brussels, Belgium\relax                                                                                                                                            
   	\label{inst:10}
   	\and Leibniz Institute for Astrophysics Potsdam (AIP), An der Sternwarte 16, D-14482 Potsdam, Germany\relax                                                                                                  
   	\label{inst:11}
   	\and Laboratoire Univers et Particules de Montpellier, Universit\'{e} Montpellier, CNRS, Place Eug\`{e}ne Bataillon, CC72, F-34095 Montpellier Cedex 05, France\relax                                        
   	\label{inst:12}
   	\and Astronomical Institute, Faculty of Mathematics and Physics, Charles University, V Hole\v{s}ovi\v{c}k\'{a}ch 2, 180 00 Prague, Czech Republic
   	\label{inst:12bis}
   	\and Laboratoire d'astrophysique de Bordeaux, Universit\'{e} de Bordeaux, CNRS, B18N, all{\'e}e Geoffroy Saint-Hilaire, F-33615 Pessac, France\relax	\label{inst:13}
   	\and Sorbonne Universit\'e CNRS, UMR 7095, Institut d'Astrophysique de Paris, 75014 Paris, France
   	\label{inst:17}
   	\and Faculty of Mathematics and Physics, University of Ljubljana, Jadranska ulica 19, SLO-1000 Ljubljana, Slovenia\relax                             
   	\label{inst:14}
   	\and Institute for Computational Cosmology, Department of Physics, Durham University, Durham DH1 3LE, UK
   	\label{inst:0018}   	
   	\and European Southern Observatory, Karl-Schwarzschild-Str. 2, 85748 Garching-bei-M\"unchen, Germany
   	\label{inst:0019}   	
             }

   \date{Accepted by A\&A June 23, 2022;}

 
  \abstract
   {The third release of the \Gaia\ catalogue contains the radial velocities for 33\,812\,183 stars having effective temperatures ranging from 3100 K to 14\,500 K. The measurements are based on the comparison of the observed RVS spectrum (wavelength coverage: 846--870\,nm, median resolving power: 11\,500) to synthetic data broadened to the adequate Along-Scan Line Spread Function. The additional line-broadening, fitted as it would only be due to axial rotation, is also produced by the pipeline and is available in the catalogue (field name \vbroad). }
   {To describe the properties of the line-broadening information extracted from the RVS and published in the catalogue, as well as to analyse the limitations imposed by the adopted method, wavelength range, and instrument.}
   {We use simulations to express the link existing between the line broadening measurement provided in \Gaia\ Data Release 3 and \vsini. We then compare the observed values to the measurements published by various catalogues and surveys (GALAH, APOGEE, LAMOST, ...).
   }
   {While we recommend being cautious in the interpretation of the \vbroad\ measurement, we also find a reasonable global agreement between the \Gaia\ Data Release 3 line broadening values and those found in the other catalogues. 
   We discuss and establish the validity domain of the published \vbroad\ values. The estimate tends to be overestimated at the lower \vsini\ end, and at $\teff>7500\,\mathrm{K}$ its quality and significance degrade rapidly when $\grvs>10$. Despite all the known and reported limitations, the \Gaia\ Data Release 3 line broadening catalogue contains the measurements obtained for 3\,524\,677 stars with \teff\ ranging from 3500 to 14\,500 K, and $\grvs<12$. It gathers the largest stellar sample ever considered for the purpose, \corr{and allows a first mapping of the \Gaia\ line broadening parameter across the HR diagram}.}
   {}

   \keywords{Stars: rotation -- Catalogs -- Techniques: spectroscopic
               }

   \maketitle
%

\section{Introduction}

\corr{In addition to its high quality astrometry, the ESA \Gaia\ space mission  provides valuable spectroscopic data. The satellite has on board an intermediate resolving power spectrometer that covers the 846 to 870\,nm wavelength range, with the initial primary goal to measure the radial velocity (RV) of the sources transiting through one of its four CCD rows \citep{CU6documentation,DR2-DPACP-46} down to the magnitude $\grvs=16.2$ \citep{DR3-DPACP-159}. During one such transit, the instrument acquires three spectra (i.e. one per CCD strip) in $\sim 4.4$\,s each. A spectroscopic pipeline processes the data \citep{2018A&A...616A...6S} to calibrate and extract the transit spectra, then derives the RV, as well as a line broadening parameter, through the Single Transit Analysis (STA) and Multiple Transit Analysis chains (MTA).}
The third release of the \Gaia\ catalogue contains the  radial velocity of 33\,812\,183 stars with effective temperatures ranging from 3100 to 14\,500\,K. Its measurement is based on the comparison of observed to synthetic template spectra \citep[][]{2014A&A...562A..97D}, and assumes that the central wavelength, strength and shape of the observed spectral lines are accurately known. Various physical phenomena can contribute to broadening or shifting the intrinsic profile of spectral lines. They relate to quantum mechanics, particle interaction, or to motions with velocity fields having scales shorter than the photons' mean free path. In most cases, the magnitude of their impact on the spectra is well described by classical atmosphere modelling and, usually, the spectral line shapes can be predicted by \elena{keeping} the effective temperature, surface gravity, metallicity, and microturbulence \elena{fixed}. Therefore the adopted method relies on a set of synthetic spectra libraries covering the astrophysical parameters (APs) space (\teff, \logg, [M/H]) and on the knowledge of the stars' APs \citep{DR3-DPACP-159, DR3-DPACP-151, DR3-DPACP-161}.  


For most targets, the line broadening at the median resolving power of the RVS \citep[$R=11\,500$, $\sim$$26 $\,\kps,][]{DR2-DPACP-46} is expected to be dominated by the instrumental spectroscopic line spread function \citep[Along-scan Line Spread-Function, henceforth LSF,][]{CU6documentation}. There are, however, other mechanisms which may also significantly broaden the lines and which require the measurement of extra parameters. The most significant of these is stellar axial rotation, whose line broadening is due to the Doppler effect and depends on the equatorial rotational velocity, $V$, and on the star's inclination angle, $i$. 

Rotational broadening leads to line-blending and hence to complex template mismatches that impact the RV measurements. Therefore, a first attempt to derive \vsini\ was included in the \corr{STA and MTA chains} \citep{2018A&A...616A...6S,CU6documentation}. On the other hand, it is known that phenomena other than stellar rotation may contribute to broadening the spectroscopic features (e.g. macroscopic random motions such as macroturbulence \elena{-- \vmacro} -- and large convection eddies, prominences, radial and non-radial pulsations, systematic velocity fields related with stellar winds, ignored binarity, limited accuracy of the LSF or of the straylight correction, ...). \elena{We did not try to disentangle their impact on the line profiles from the rotational broadening, and ignored these when estimating \vsini\ (e.g. the synthetic spectra adopted assumes $V_\mathrm{macro} = 0$\,\kps).}

Therefore, while the line broadening is measured with a classic rotational kernel, the measurement provided in the catalogue is named \linktoparam{gaia_source}{vbroad}. For the same reason, in what follows, \vbroad\ refers to the estimate provided by the \corr{STA/MTA parts of the spectroscopic pipeline}, while \vsini\ denotes the true projected rotational velocity value \elena{(e.g. from simulations) or the value found in other catalogues or surveys (i.e. even when the catalogue/survey itself does not disentangle \vsini\ from other broadening mechanisms, and/or similarly gives a different name to the estimate)}.








\corr{Another estimate of the RVS line broadening is obtained by the \textit{ESP-HS}\footnote{\corr{Extended Stellar Parametrizer -- Hot Stars}} module of the {\it Apsis}\footnote{\corr{Astrophysical ParameterS Inference System}} pipeline \citep{DR3-DPACP-157}. It is published in \GDRthree\ as \linktoaps{vsini_esphs} (in the {\tt astrophysical\_parameters} table) and is an intermediate result of the analysis of the RVS and BP/RP data when deriving the astrophysical parameters of stars with $\teff\,>\,7500\,K$. A discussion of {\tt vsini\_esphs} and a comparison with the \vbroad\ measurements described in the present paper is given in the online documentation 
\citep[Section~11.4.4,][]{APdocumentation}
as well as in \citet{DR3-DPACP-160}.} In this work, we aim at providing to the \GDRthree\ catalogue user more information on the line broadening parameter derived from the spectra obtained by the \Gaia\ Radial Velocity Spectrometer (RVS) \corr{and derived by the spectroscopic pipeline}. The adopted method to derive it, as well as the expected accuracy, limitations, and significance are described in Sect.\,\ref{sec:method}. We provide a general overview of the results in Sect.\,\ref{sec:results}. During the validation process, the pipeline output was compared to values found in various catalogues. We report our findings in Sect.\,\ref{sec:catalogues} and discuss the statistical behaviour, offsets and dispersion in Sect.\,\ref{sec:discussion}. Our main conclusions are summarised in Sect.\,\ref{sec:conclusions}.

\section{Method\label{sec:method}}

\subsection{Description\label{sec:method:description}}
\begin{figure}[!htb]
{\center
\includegraphics[width=0.99\columnwidth,clip=,draft=False]{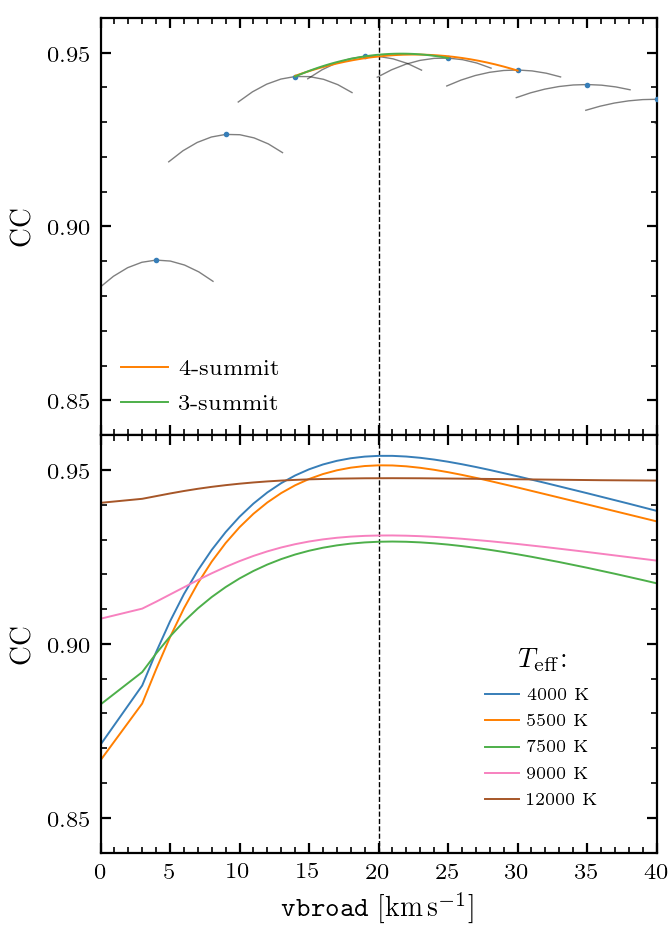}
 \caption{\vbroad\ determination at $\teff=5500$\,K, $\logg=4.5$, $\feh= 0$, $\vbroad=20$\,\kps\ \corr{(vertical dashed line)}, and $\grvs=8$. Template mismatch errors are ignored, except for the \vbroad\ broadening which is the quantity to derive. Upper panel: The top of the cross-correlation functions centred at 0\,\kps\ (grey curves) and obtained by assuming various values of \vbroad\ is plotted and shifted according to the adopted \vbroad.
  Their summit is identified by blue circles, while the 3-summit and 4-summit parabola fits are shown by green and orange curves, respectively. The ordinate axis label `CC' stands for `cross-correlation coefficient'. Lower panel: Same as in the upper panel, but at different effective temperature values. For readability reasons the CCF summits are connected by a line.}\label{fig:method}
 }
\end{figure}

The \vbroad\, determination is a part of the \corr{STA} and \corr{MTA chains} of \corr{the spectroscopic pipeline that is assuming to only be applied on single-lined spectra with no emission. Suspicion for line-emission or binarity \citep{CU6documentation, DR3-DPACP-159, DR3-DPACP-161} is usually detected by the pipeline. About 28\,000 targets were flagged for having emission in their spectra, and $\sim40\,000$ \citep[][]{DR3-DPACP-159} were flagged as being SB2 candidates by the  pipeline. Therefore, these were not processed for (single-lined) RV and \vbroad.} \corr{For all the other cases, the  measurement} is performed on a transit per transit basis by maximising the top of the combination of the cross-correlation functions (CCF) that result from the correlation of all the valid CCD strip spectra by the template which is broadened to a given \vbroad\ value \elena{(upper panel in Fig.\,\ref{fig:method})}. The `template' is the continuum-normalised and LSF-broadened synthetic spectrum whose set of APs in the library is the nearest to the target's parameters. \elena{The library of synthetic spectra we have adopted is described in \citet{RHB-005} and does not include any additional line-broadening (e.g. ignores macroturbulence).} For the stars cooler than 7000\,K, most of the parameter values were taken from intermediate results of {\it Apsis} \citep{DR3-DPACP-157} with an earlier version of {\it GSP-Phot}\footnote{\corr{General Stellar Parametrizer from Photometry}} and of {\it GSP-Spec}\footnote{\corr{General Stellar Parametrizer from Spectroscopy}} \citep[][note that these papers describe the results obtained with \GDRthree\  BP/RP and RVS spectra]{DR3-DPACP-156, DR3-DPACP-186}, as well as with DR2 spectra, while for the hotter ones they were derived as explained by \citet{DR3-DPACP-151} to reduce the impact of known mismatches on the RV determination \citep[see Sect.\,6.4.8 of][for more information on the STA pipeline and the determination of \vbroad.]{CU6documentation}. Furthermore, during the pipeline testing/validation process that preceded the operational run, no time was left to assess the impact of de-blended spectra on the measurement of \vbroad, therefore it has been decided to remain conservative and to derive it using non-blended spectra only.

\elena{The CCF maximisation procedure allows \vbroad\ to vary, in three iterations from 0 to 600 \kps\ (i.e. each iteration reducing the step around the maximum), with a minimum \vbroad\ step of 5 \kps. For each transit, the final result is obtained by adopting \corr{the procedure  described in \citet{1995A&AS..111..183D} to mitigate the impact of discretisation. As shown in Fig.\,\ref{fig:method} (upper panel), the approach combines the solution obtained by fitting two parabolas through 3- and 4-points (see their Eq.\,19) taken around the top of the function defined by the CCF maxima estimated at different \vbroad\ values. Hence, we assumed that the top of the function to fit is nearly symmetrical. In practice, the existing asymmetry makes the procedure less effective but still meaningful in most cases.}

We show in the lower panel of the same figure how the sensitivity of the CCF maximisation varies with the effective temperature and the spectroscopic content of the RVS. The dependence of the CCF maxima with \vbroad\ is stronger at lower values and flattens with increasing \teff, especially above 7500 K.} While one estimate per transit is determined (at the STA stage), the target \vbroad\, that is published in the \GDRthree\ catalogue is the median taken (during the MTA stage) over at least six valid transits (Sect.\,\ref{sec:method:filters}), and the corresponding uncertainty is assumed to be equal to the standard deviation.

\subsection{Post-Processing filtering \label{sec:method:filters}}

In the present paper, we report on the \vbroad\ estimates published in the \GDRthree\ catalogue.
\corr{Prior to post-processing, 7\,218\,658 \vbroad\ estimates were available for sources with $\grvs \le 12$.} About 50 percent of the results initially available were filtered out \corr{after quality assessment}. We established the filtering criteria during the validation of the pipeline results as follows:

\begin{enumerate}
\item Most \vbroad\, values and uncertainties of targets with less than six transits showed dubious features, and were therefore removed from the catalogue \corr{(i.e. keep value when $N_\mathrm{t} \ge 6$)}.
\item Because the rotational convolution is performed in Fourier space, with a sampling of the spectra that was optimised for RV determination ($\sim$4\,\kps), all values lower than 4\,\kps\ are questionable. For this reason, we filtered out all estimates less than or equal to 5\,\kps\ \corr{(i.e. keep when $\vbroad > 5\,\kps$)}.
\item \vbroad\ values higher than 500 \kps\ were removed as they formed a noticeable and likely non-physical overdensity in the observed velocity distribution \corr{(i.e. keep when $\vbroad < 500\,\kps$)}.
\item In the very cool temperature range and in the valid \vbroad\, domain, we found too few catalogue values to validate the  measurements. It was therefore decided to filter out the estimates obtained for stars cooler than 3500\,K \corr{(i.e. keep when $\teff \ge 3500\,\mathrm{K}$)}.
\item For consistency reasons, \vbroad\ measurements obtained on data having no valid radial velocity were deleted \corr{(i.e. keep when RV is valid)}. Therefore, with the previous filter taken into account, only measurements obtained for targets with \teff\ ranging from 3500\,K to 14\,500\,K are published. 
\end{enumerate}


\subsection{Expected accuracy and significance\label{sec:method:significance}}

The Radial Velocity Spectrometer covers the 846--870\,nm wavelength domain \citep{DR2-DPACP-46}, with a median resolving power of 11\,500. The selection of the wavelength domain is a compromise between technical and astrophysical constraints. The goal being to measure the most accurate radial velocities for the majority of the stellar populations seen by \Gaia\ with the most accurate astrometry. The calcium triplet observed in this domain was found to be the best choice \citep[e.g.][]{1999BaltA...8...73M}, as it remains strong at various metallicity regimes in the spectra of F-, G-, and K-type stars. 

\begin{figure}[!htb]
{\center
\includegraphics[width=0.99\columnwidth,clip=,draft=False]{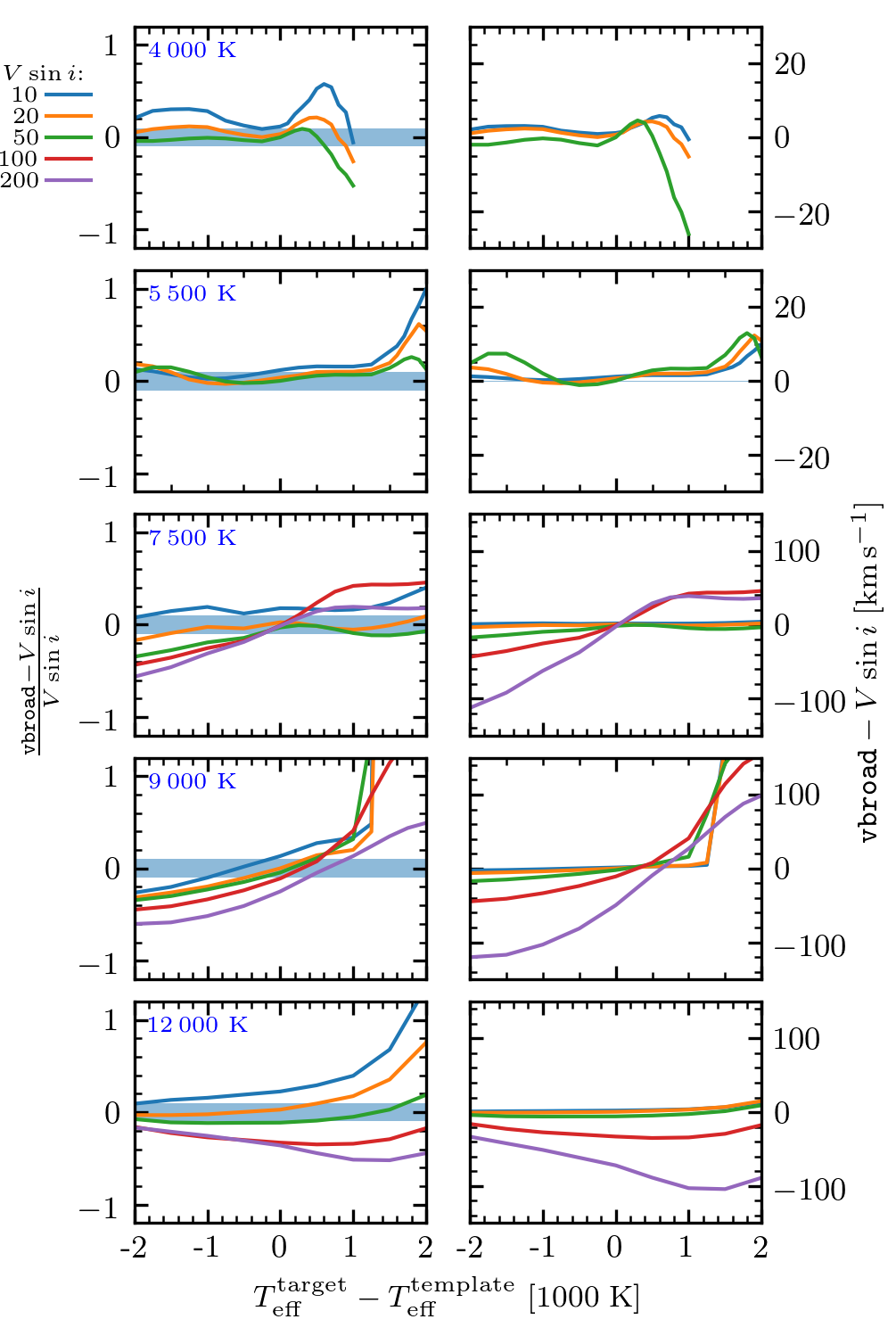}
 \caption{Relative (left panels) and absolute (right panels) \vbroad\ $-$ \vsini\ residuals plotted as a function of the \teff\ error made during the selection of the template spectrum. \vsini\ stands for the projected rotational velocity adopted to construct the simulation, while \vbroad\ is the estimate provided by the pipeline. Different \vsini\ (see legend and colour  coding) and `true' \teff\ estimates are considered. In the left panels, the blue hatches identify the domain where the errors are within 10\% of the expected value.}\label{fig:template_mismatch}
 }
\end{figure}

While rotational broadening may have an impact on the RV determination, the RVS domain is not well suited for its accurate determination. This is especially the case for stars hotter than 7000\,K where the main features are due to intrinsically broad lines (higher members of the Paschen series and \ion{Ca}{ii} triplet lines), which by nature are strongly blended with one another \citep[e.g. Fig.\,17 in][]{DR2-DPACP-46}. Further, with the adopted methodology, the measurement of \vsini\, or \vbroad\, strongly depends on the quality of the template spectrum, which in turn supposes a good knowledge of the astrophysical parameters and of the phenomena that shape the line profiles. Consequently, a wrong template will automatically lead to an incorrect estimate.

To test the impact of \teff\ template mismatch by ignoring noise and assuming a perfect knowledge of the LSF (i.e. for the exercise we assumed a Gaussian LSF and a resolving power of 11\,500), we ran a partial version of the pipeline that derives \vbroad\ on synthetic spectra and chose, for the same target spectrum, templates with various \teff\ mismatches/errors. Figure\,\ref{fig:template_mismatch} shows the results obtained at different \vsini\ and effective temperature values on the main sequence (MS). We extend the explored range of \teff\ mismatches up to $\pm2000$\,K to cover most of the possible cases, but one usually expects lower errors/mismatches especially for the late\elena{-type} stars.
The impact of the template mismatch depends on the sign and absolute value of the \teff\ error. In most cases, the \vbroad\ estimate is more sensitive to positive temperature errors (i.e. the template \teff\ is lower than the target \teff) when the template usually exhibits more spectral features. In these cases, the pipeline tends to overestimate \vbroad. On the other hand, in A-type stars, where the blends between the Paschen and calcium triplet lines dominate, the accuracy of \vbroad\ is the most sensitive to the \teff\ error. A similar negative impact of the \teff\ error on the RV estimates of the A-type stars has also been noted \citep{DR2-DPACP-54}, and led to the redetermination of the APs \citep[Sect.\,3 of][]{DR3-DPACP-151}, as well as to a first estimate of the line broadening,  by the pipeline before RV and \vbroad\ are derived. For that reason, as the same template is used for RV and \vbroad\ determination, the effect of template mismatch due to inaccurate APs is expected to be mitigated for the A- and B-type stars. 

\begin{figure}[!htb]
{\center
\includegraphics[width=0.89\columnwidth,clip=,draft=False]{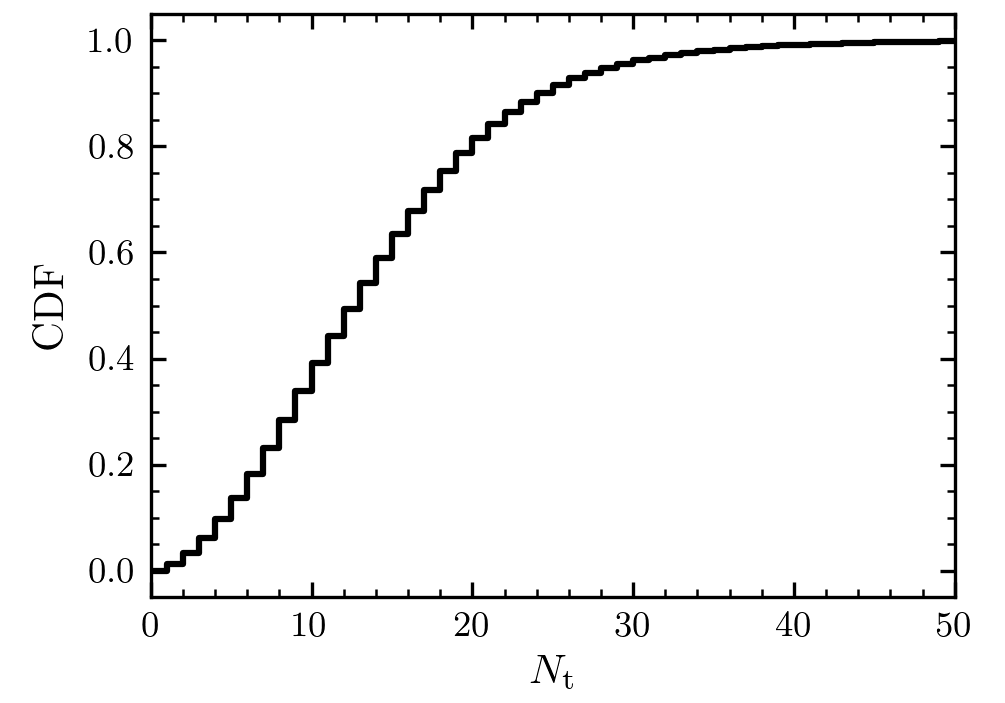}
 \caption{Cumulative distribution function of the number of `not blended' transits ($N_\mathrm{t}$) before post-processing.}\label{fig:nt_cdf}
 }
\end{figure}

Furthermore, we conducted a series of Monte-Carlo (MC) simulations which aimed to better illustrate the limitations of the technique/pipeline and of the wavelength domain adopted, as well as of the instrument (in particular its resolving power). One MC sample is made up of 1000 cases at a fixed \teff, \logg, [M/H], and \grvs\ magnitude. Each of these MC realisations assumes a different number of transits ($N_\mathrm{t}$), and \vsini, while each CCD strip spectrum has its own photon noise. The number of transits is randomly chosen but follows the observed $N_\mathrm{t}$ distribution (Fig.\,\ref{fig:nt_cdf}), while \vsini\ ranges from 0 to 600 \kps\ and follows a uniform random distribution. No template mismatch was introduced during the tests, and the same post-processing filters were applied (e.g. only cases with more than five transits are considered, see Sect.\,\ref{sec:method:filters}).

\begin{figure}[!htb]
{\center
\includegraphics[width=1\linewidth,clip=,draft=False]{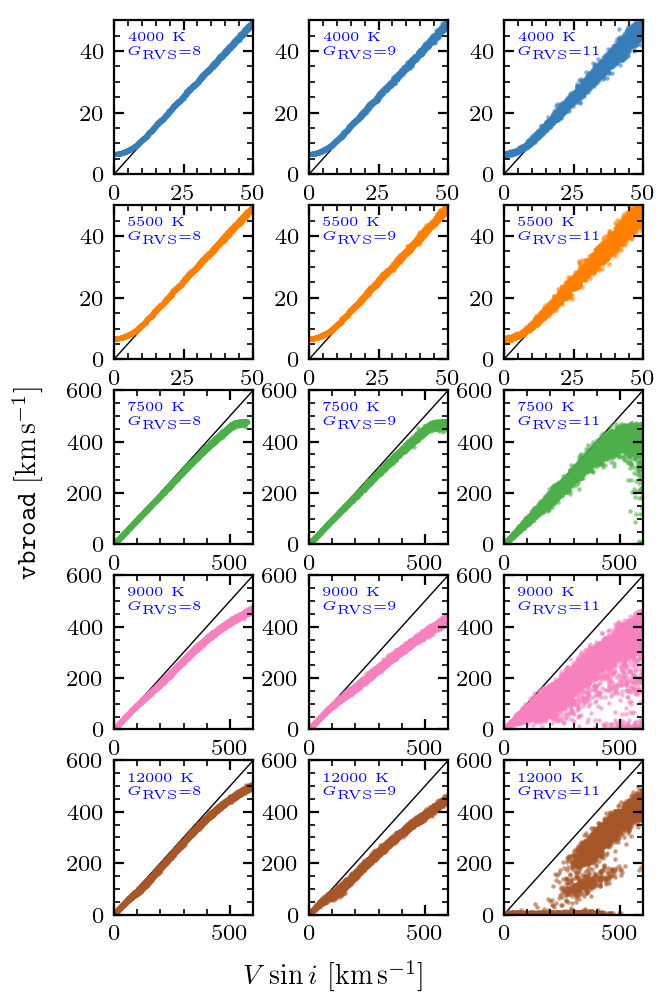}
 \caption{Monte-Carlo simulations:  \vbroad\ as a function of \vsini\ for various \grvs\ magnitudes and effective temperatures. The identity relation is represented by the black line. The colour  coding is the same as in Fig.\,\ref{fig:mc.rel.error.ms.sol}.}\label{fig:mc.abs.error.ms.sol}
 }
\end{figure}

\begin{figure}[!htb]
{\center
\includegraphics[width=0.99\columnwidth,clip=,draft=False]{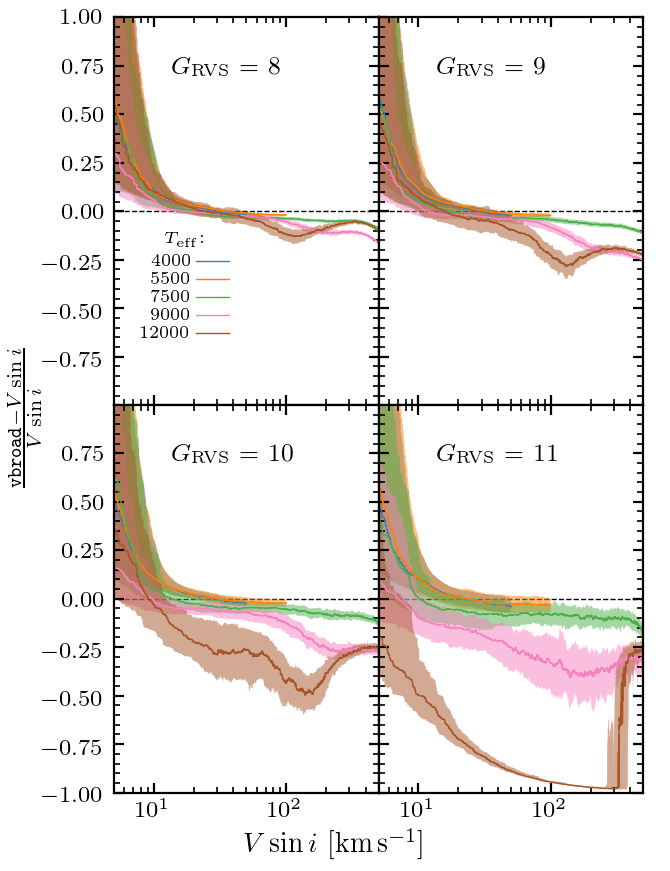}
 \caption{Monte-Carlo simulations: Relative (\vbroad\ $-$ \vsini) residuals as a function of \vsini\ for various \grvs\ magnitudes and effective temperatures (coloured lines). The 15--85\% interquantile range is represented by shades.}\label{fig:mc.rel.error.ms.sol}
 }
\end{figure}

\begin{figure*}[!htb]
\sidecaption
\includegraphics[width=12cm,clip=,draft=False]{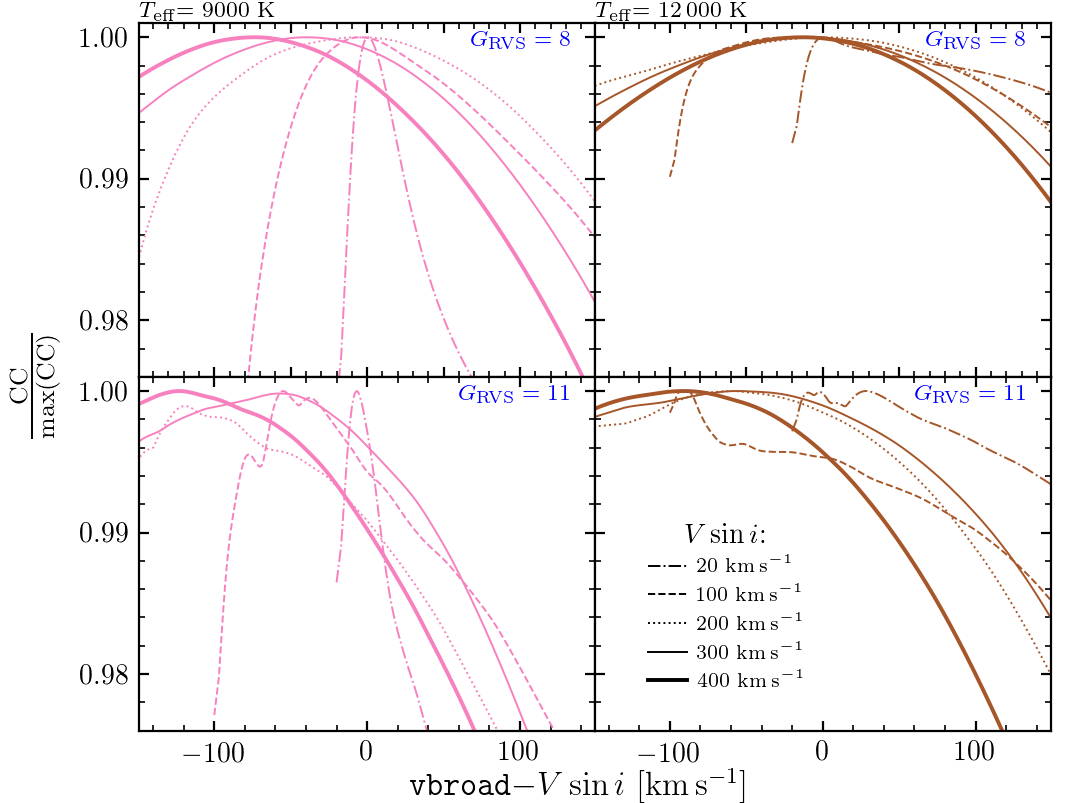}
 \caption{Example of variation of the CCF maximum with \teff, \grvs\ (noted in blue in the upper right corner of each panel), \vbroad, and \vsini\ (see line styles in the legend). Each curve, which represents only one noise realisation (i.e. one transit), is normalised to its highest value at a given \vsini. See also Fig.\,\ref{fig:method}.
 }\label{fig:CCF}
\end{figure*}

The main outcomes of the tests are illustrated in Fig.\,\ref{fig:mc.abs.error.ms.sol} where \vbroad\ is plotted as a function of \vsini, and in Fig.\,\ref{fig:mc.rel.error.ms.sol} which shows how the relative error varies with \vsini. Both figures were made for different combinations of the effective temperature and magnitude. At \vsini\ lower than 20\,\kps, due to the resolving power, the wavelength sampling and the approach we adopted, \vbroad\ tends to be systematically larger than \vsini. At higher values, the error remains within 10\% for the brightest magnitudes with a \vbroad\ measurement that tends to be underestimated. When the magnitude gets fainter, the results degrade rapidly at $\teff>7500$\,K. In temperature regime of the early A- and B-type stars, the impact of the broadening on the Paschen lines remains the main source of information available. We show in Fig.\,\ref{fig:CCF}, for one transit and one noise realisation, how the CCF maximum varies with \vbroad, \vsini, \grvs, and \teff\ above 7500\,K. At 9000\,K, where the Paschen lines are the largest/broadest and blended with the calcium triplet, the offset strongly increases with \vsini\ (Fig.\,\ref{fig:CCF}, upper left panel). The CCF centre is most sensitive (i.e. its gradient with \vbroad\ varies more rapidly) at lower \vsini\ for $\grvs=8$, but it rapidly becomes noisier with increasing magnitude (Fig.\,\ref{fig:CCF}, lower left panel). Conversely, at 12\,000\,K, and with a spectrum dominated by the overlapping Paschen lines, the method tends to be less sensitive to low \vsini\ (i.e. smaller curvature, see right upper panel of Fig.\,\ref{fig:CCF}), and still decreases rapidly with magnitude (Fig.\,\ref{fig:CCF}, right lower panel).
These effects, as well as the limitations  inherent to our measuring technique, are at the origin of the features seen at low \vsini\ in the lower right panel of Figs.\,\ref{fig:mc.rel.error.ms.sol} and \ref{fig:mc.abs.error.ms.sol} ($\teff=12\,000$\,K, $\grvs=11$).


\section{Results\label{sec:results}}

The post-processed results of the \vbroad\ determination algorithm are to be found in the {\tt gaia\_source} table. Fields \linktoparam{gaia_source}{vbroad}  and \linktoparam{gaia_source}{vbroad_error}  contain the \vbroad\ estimate and its standard deviation, respectively. The number of transits considered to compute the median is given in \linktoparam{gaia_source}{vbroad_nb_transits}.

\begin{table}[]
    \centering
        \caption{Impact of the post-processing on the number of remaining, $N_\mathrm{rem.}$, \vbroad\ estimates. $N_\mathrm{rem.}$ takes into account all the filters previously applied (i.e. current and previous table rows). The filters are listed with their item number (\#) from Sect.\,\ref{sec:method:filters}.}
    \label{tab:sample.budget}
    \begin{tabular}{ccc}
    \hline
    \hline
        \# & Filter & $N_\mathrm{rem.}$ \\
        \hline
        & $\grvs \le 12$ & 7\,218\,658 \\
        1 & $N_\mathrm{t} \ge 6$ & 5\,327\,091 \\
        2 & $\vbroad > 5\,\kps$ & 3\,717\,427 \\
        3 & $\vbroad < 500\,\kps$ & 3\,717\,143\\
        4 & $\teff \ge 3500\,\mathrm{K}$ & 3\,675\,448 \\
        5 & RV is valid & 3\,524\,677\\
    \hline
    \end{tabular}
\end{table}

\begin{figure}[!htb]
{\center
\includegraphics[width=0.99\columnwidth,clip=,draft=False]{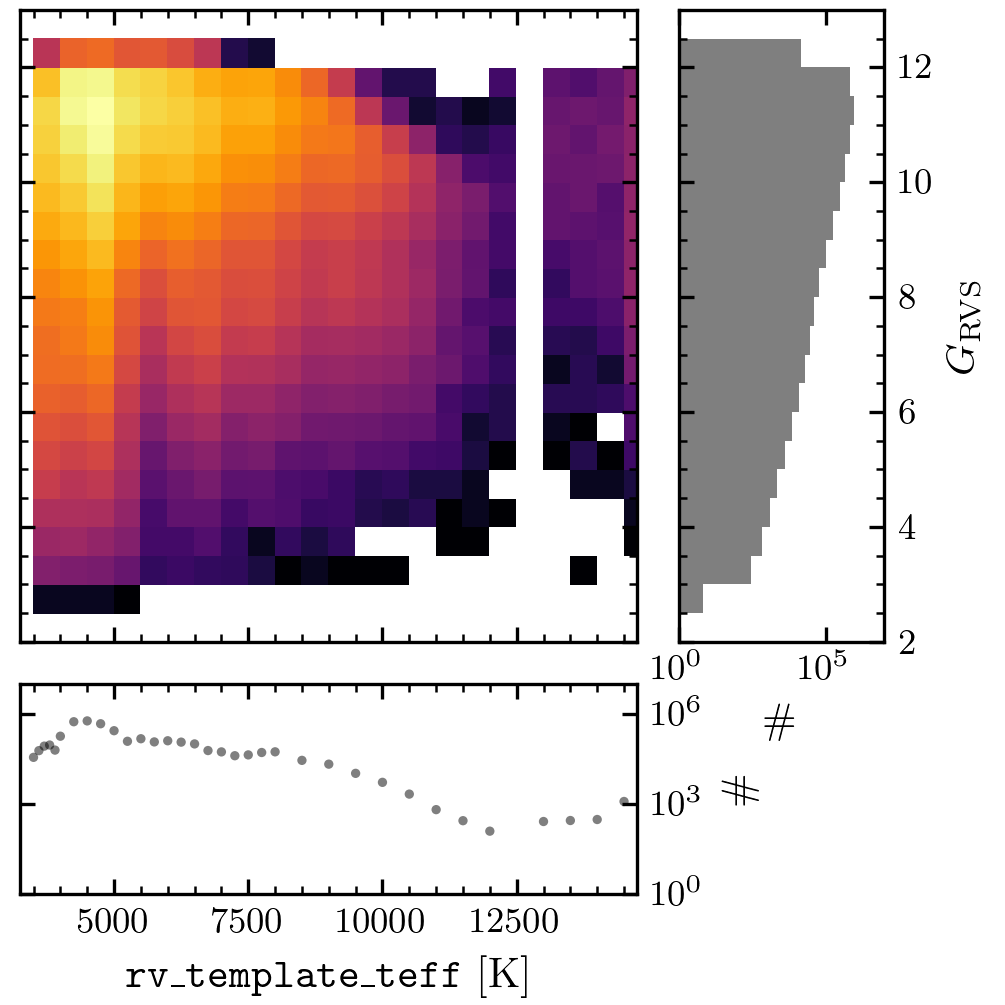}
 \caption{Distribution of the \GDRthree\ \vbroad\ catalogue with magnitude and effective temperature. Lower panel: Effective temperature of the adopted template (\linktoparam{gaia_source}{rv_template_teff}) distribution. Note that our
library of templates does not have any spectra with \teff = 12\,500 K, which translates into a gap in the distribution at the same temperature. Right panel: \grvs\ magnitude (\linktoparam{gaia_source}{grvs_mag}) distribution.}\label{fig:teff.mag.dist}
 }
\end{figure}

\begin{figure}[!htb]
{\center
\includegraphics[width=0.99\columnwidth,clip=,draft=False]{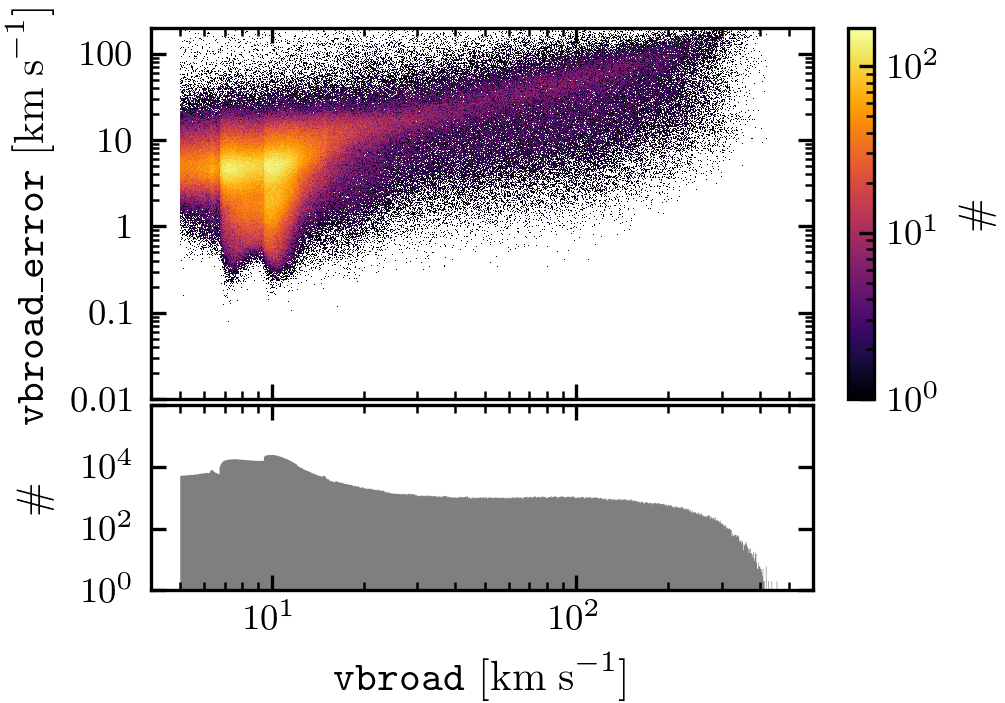}
 \caption{Upper panel: {\tt vbroad\_error} vs. {\tt vbroad}. Lower panel: Corresponding distribution of the number of targets in each {\tt vbroad} bin.\label{fig:fish.after}}
 }
\end{figure}

\corr{The impact of the successive post-processing filters (Sect\,\ref{sec:method:filters}) is summarised in Table\,\ref{tab:sample.budget}. From the 7\,218\,658 \vbroad\ measurements initially available for targets brighter than $\grvs=12$, 3\,524\,677 are published in the \GDRthree\ catalogue. Their magnitude and \teff\ distributions are given in Fig.\,\ref{fig:teff.mag.dist}.}
Among these, 428\,529\footnote{The number of available spectra was obtained by forming the following query: {\tt SELECT * FROM user\_dr3int6.gaia\_source WHERE vbroad is not null and has\_rvs ='t'}} stars have their spectra published with, however, an expected resolution lower than the CCD spectra \citep{DR3-DPACP-154}. As a consequence of the post-processing (Sect.\,\ref{sec:method:filters}), the adopted template \teff\ ranges from 3500\,K to 14\,500\,K. No measurement is expected for stars fainter than the magnitude 12. However,  during the processing the decision is based on a \grvs\ estimate slightly different from the one published in the field \linktoparam{gaia_source}{grvs_mag} \citep{DR3-DPACP-155} which is plotted in Fig.\,\ref{fig:teff.mag.dist} and explains that a fraction of fainter targets are present. 
The variation of {\tt vbroad\_error} with \vbroad\ is represented in Fig.\,\ref{fig:fish.after}. The stellar population of \Gaia\ is dominated by slowly rotating FGK stars, which produces the overdensity at \vbroad\ $<$ 20 \kps. 

\begin{figure}[!htb]
{\center
\includegraphics[width=0.99\columnwidth,clip=,draft=False]{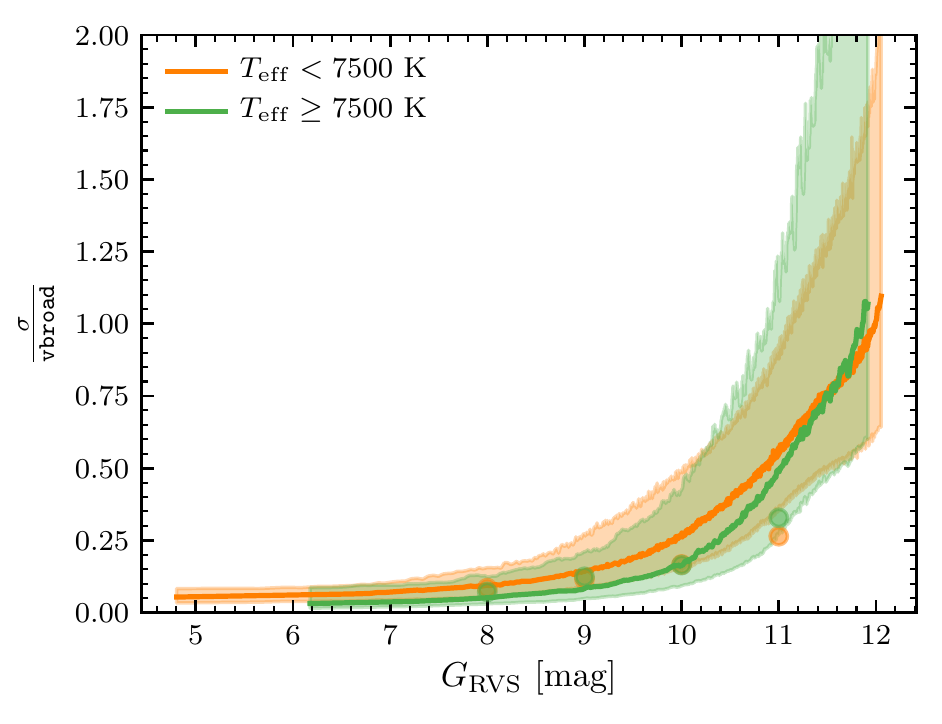}
 \caption{Relative uncertainty on \vbroad\ as a function of \grvs\ magnitude, for two \teff\ ranges. Thick lines are the running median values (over 2000 targets) and the coloured regions correspond to the associated 15\% and 85\% quantiles. The filled circles are the relative uncertainties corrected from the z-score estimations performed in Sect.\,\ref{sec:discussion}. \label{fig:relsigvsGrvs}}
 }
\end{figure}


Figure\,\ref{fig:relsigvsGrvs} displays the variation of the relative uncertainty $\frac{\mathtt{vbroad\_error}}{\vbroad}$ as a function of \grvs\ magnitude for cool ($\teff<7500$\,K) and hot stars ($\teff\geq 7500$\,K). The relative uncertainty remains better than 20\% for targets brighter than $\grvs=9$, but significantly increases for fainter objects: it reaches 60\% at $\grvs=11$ until it exceeds 100\% at the faint limit. 

\section{Comparison with other catalogues and surveys\label{sec:catalogues}}

\begin{figure}[!h]
{\center
\includegraphics[width=0.99\linewidth,clip=,draft=False]{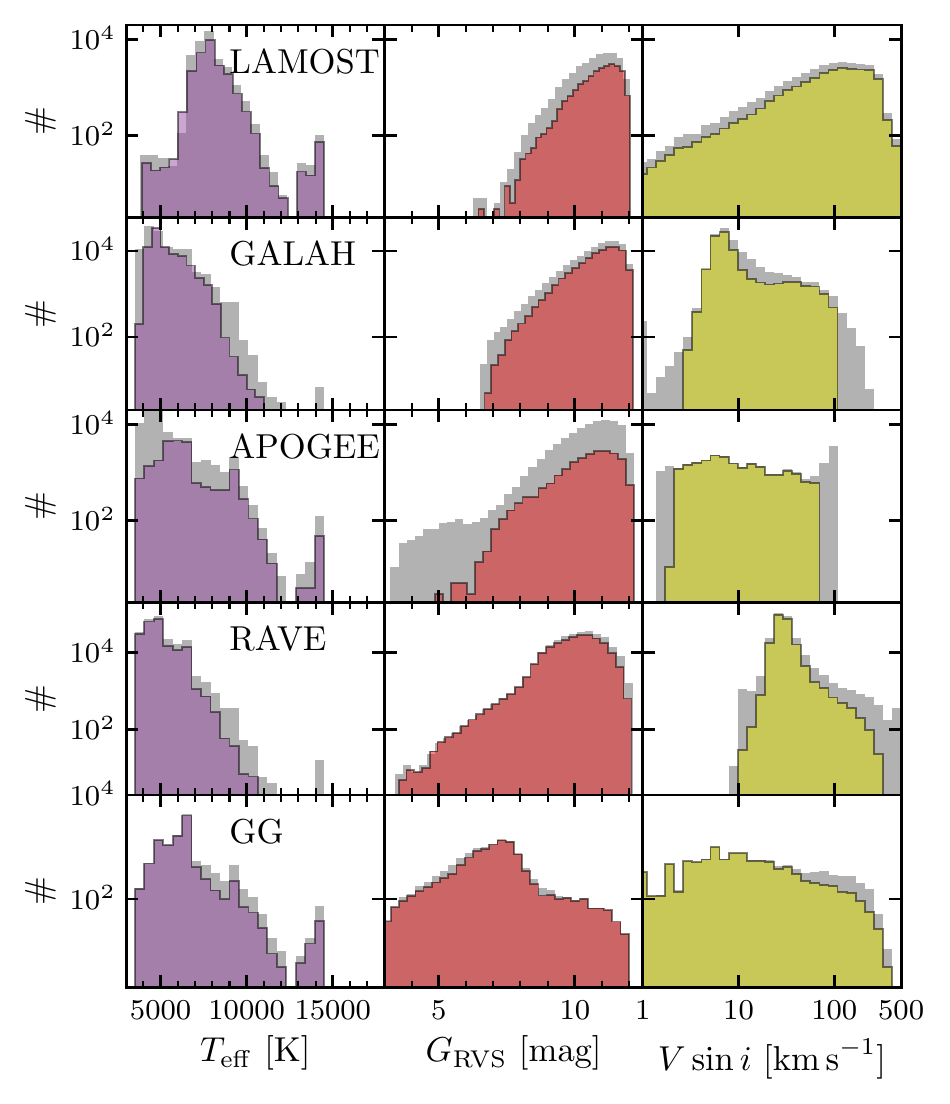}

 \caption{Overview of the intersection of stars having published \GDRthree\ \vbroad\ with the reference catalogues/surveys: GG \citep{Glebocki}, RAVE \citep{RaveDR6}, GALAH \citep{GalahDR3}, APOGEE \citep{ApogeeDR16}, LAMOST \citep{LamostDR6}. Left panel: \teff\ (i.e. {\tt rv\_template\_teff}) distribution. Central panel: \grvs\ distribution. Right panel: \vsini\ distribution. The additional grey bins represent the part of catalogue intersections discarded in the cleaning process (see Sect.\,\ref{sec:catalogues:selection}).\label{fig:cats.overview}}
 }
\end{figure}

The large spectroscopic surveys that have been initiated in the last two decades have published a huge quantity of rotational broadening measurements. These homogeneous sets of values provide a way to compare the different scales of rotational broadening measurements, each of them being affected by their own biases and uncertainties, originating from determination methods or from  instrumental configuration.
Four different catalogues have been chosen to compare the \GDRthree\ \vbroad\ parameters with: RAVE DR6 \citep{RaveDR6}, GALAH DR3 \citep{GalahDR3}, APOGEE DR16 \citep{ApogeeDR16}, LAMOST DR6 (OBA stars) \citep{LamostDR6}. In addition to these, the compilation made by \citet[hereafter referred to as GG]{Glebocki} allows a comparison for \vbroad\ values determined on brighter targets. An overview of the catalogues and surveys we have considered is given in Fig.\,\ref{fig:cats.overview}, and shows the coverage in terms of \teff, \grvs\ and \vsini\ for the different comparison samples. The spectral characteristics of the catalogues and the size of the comparison samples are summarised in Table\,\ref{tab:catalogues}.

\begin{table}[!h]
    \centering
        \caption{Characteristics of the comparison catalogues. The size is the one of the comparison sample.}
    \label{tab:catalogues}
    \begin{tabular}{lrrr@{~}c@{~}r}
         \hline
         \hline
       Catalogue &  \multicolumn{1}{c}{Size} & Resolution & \multicolumn{3}{c}{Spectral range} \\
                 &       &           &   \multicolumn{3}{c}{[nm]}\\
        \hline
         GG & 10\,821 & various & \multicolumn{3}{c}{various} \\
         RAVE & 212\,622 & 7500 & 841.0&--&879.5 \\
         APOGEE & 21\,078 & 22\,500 & 1514.0&--&1694.0 \\
         GALAH & 84\,464 & 28\,000 & 471.3&--&490.3 \\
                             &&& 564.8&--&587.3\\
                             &&& 647.8&--&673.7\\
                             &&& 758.5&--&788.7 \\
         LAMOST & 25\,770 & 1800 & 380.0&--&900.0 \\
                 \hline
    \end{tabular}
\end{table}

The RAVE pipeline operations are described in RAVE DR2 \citep{RaveDR2} and DR3 \citep{RaveDR3} papers. 
To derive the stellar parameters, they use a penalised $\chi^2$ technique to model the observed spectrum as a weighted sum of template spectra with known parameters. Due to the low spectral resolution (Table\,\ref{tab:catalogues}) and the resulting difficulty to measure low rotational velocities, they chose to restrict the dimension of their grid of templates in \vsini. Their synthetic spectra library is hence poorly populated in the low end of rotational broadening: their low \vsini\ values are only 10, 30 and 50\,\kps. \elena{The macroturbulence velocity is not part of the atmospheric parameters taken into account in the RAVE pipeline.}

For LAMOST, \citet{LamostDR6} analysed the low-resolution survey spectra of hot stars, specifically OBA, and they adapted \textit{The Payne} neural network spectral modelling method to hot stars to determine the stellar labels of the sample targets. \elena{At the resolution of LAMOST, they are not able to disentangle macroturbulence from rotational velocities, and their \vsini\ estimates include its contribution.}

 In the APOGEE pipeline \citep{ASPCAP}, the spectral analysis is performed with \textit{FERRE} \citep{FERRE}, which finds the best-fitting stellar parameters describing an observed spectrum by interpolating in a grid of synthetic templates. This grid is, however, restricted in the \vsini\ dimension to the values 1.5, 3, 6, 12, 24, 48 and 96\,\kps. \elena{\vsini\ is only determined for dwarf stars, while in the giant sub-grids a macroturbulence velocity broadening, calibrated as a function of metallicity \citep{ApogeeDR16}, is adopted instead.}

In GALAH, the stellar atmospheric parameters are derived using the spectrum synthesis code \textit{Spectroscopy Made Easy} \citep{SME}. \elena{In the corresponding catalogue, the \vsini\ parameter is cautiously named $v_\mathrm{broad}$ as it is fitted by setting the macroturbulence to 0, macroturbulent and rotational broadening influences being degenerate at the resolution of GALAH \citep{GalahDR3}.}

We used the mean \vsini\ determinations given by \cite{Glebocki}. The main contributions come from \cite{Nordstrom} providing about 12\,500 \vsini\ determined by cross-correlation technique for F- and G-dwarf stars, notably complemented by almost 3000 \vsini\ derived from FWHM for B- and A-type stars \citep{Abt02, Abt95}.
 It is worth noticing that the catalogue built by \citeauthor{Glebocki} partly inherits the discretisation of \vsini\ from the publications it compiles. This discretisation can produce an overestimation of the residuals for low \vsini\ values.

\begin{figure*}[!htp]
{\center
\includegraphics[width=0.99\linewidth,clip=,draft=False]{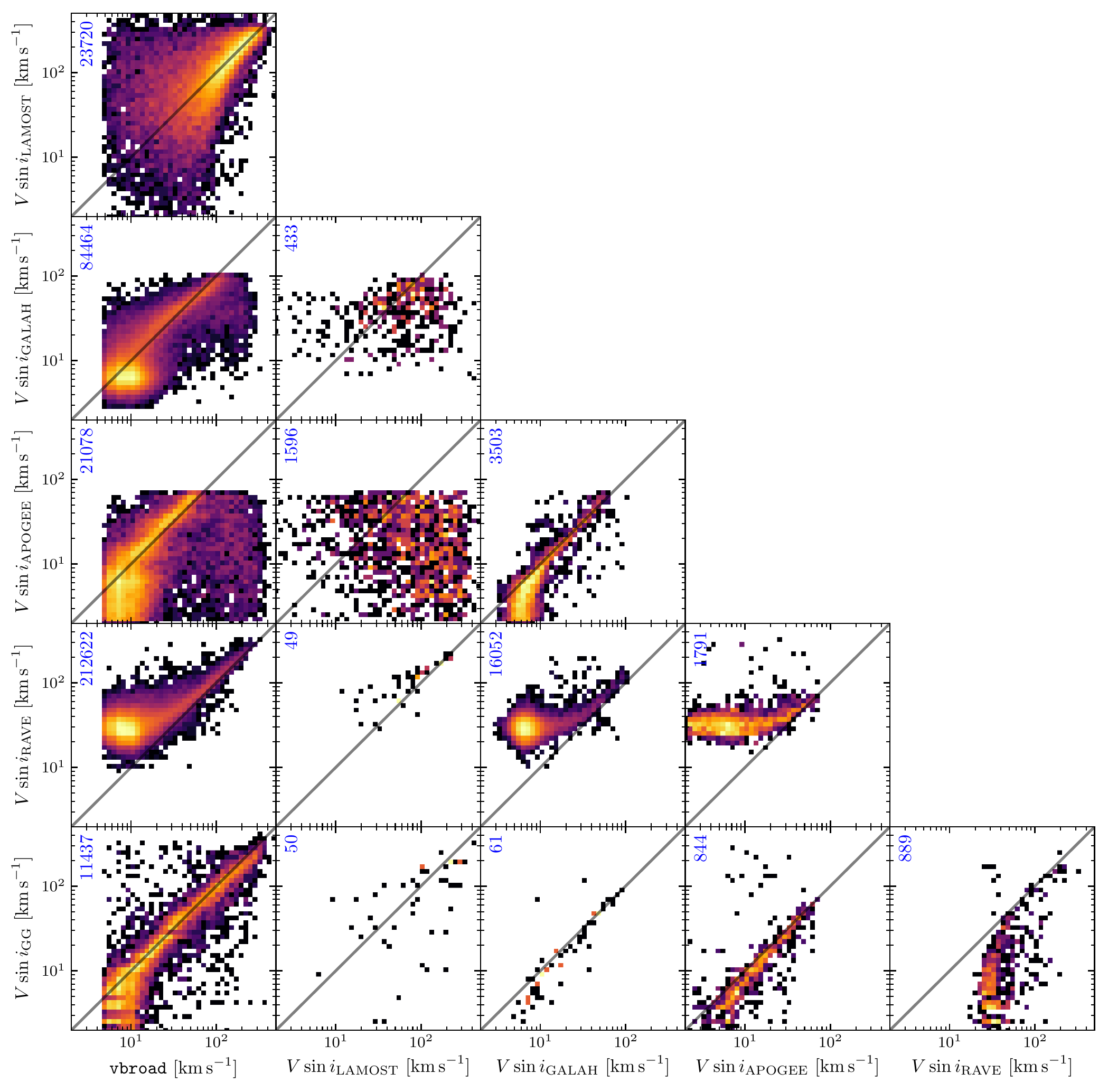}
 \caption{Comparison with other catalogues:  one-to-one comparisons of line broadening measurements between the considered sources, including \GDRthree. The velocity scales are logarithmic, as well as the density colour  scale. 
 Sizes of comparison samples are indicated in the upper left corners, while the one-to-one relation is represented by the black diagonal line.}\label{fig:compvsini}
 }
\end{figure*}

\subsection{Selection of the comparison samples\label{sec:catalogues:selection}}


The catalogues we used for comparing the line broadening scales provide, in some cases, a quality assessment of their data. We used these assessments to only keep the most trustable estimates as follows: 

\begin{itemize}
\item In the GALAH survey, the flag \texttt{flag\_sp} reflects the quality of the spectroscopic parameters, and only common targets with \texttt{flag\_sp}$ =0$ have been taken into account. 
\item The APOGEE catalogue also provides a flag, \texttt{f\_Vsini}, assessing the quality of the published \vsini\ determinations. Only common targets with \texttt{f\_Vsini}$\,=0$ are considered here. RAVE data have been selected on the basis of the height of the cross-correlation function given in the catalogue: \texttt{hcp}$\,>0.9$. 
\item LAMOST data have been selected on the basis of their reduced $\chi^2$ such that: \texttt{CHISQ\_RED}$ \,<5$. 
\item The quality of the compiled data from GG is assessed upon the flag \texttt{n\_vsini} that indicates when the precision is poor (`:') or when the datum solely originates from \citeauthor{UF}'s compilation (\citeyear{UF})\elena{, whose quality was already questioned by \citeauthor{Soderblom89} back in \citeyear{Soderblom89}.} Only targets with empty \texttt{n\_vsini} flag have been used. 
\end{itemize}

\noindent Figure\,\ref{fig:cats.overview} shows as grey bars the data that have been discarded from the comparison samples using the aforementioned criteria, and one can notice the cuts produced by this selection in the \vsini\ distributions: all targets with $\vsini\gtrsim70$\,\kps\ and $\vsini\gtrsim100$\,\kps\ are removed from the APOGEE and GALAH comparison samples, respectively.


\subsection{Two-by-two comparisons\label{sec:2by2}}

Figure\,\ref{fig:compvsini} displays the two-by-two comparisons we have made with the catalogues. The five panels on the left confront the \GDRthree\ \vbroad\ to the ground-based measurements, while the remaining panels show internal cross-matches between the catalogues, without restricting the comparison to the intersection with the \GDRthree\ values. As the GG compilation mainly contains bright targets, its intersection with the other ground-based surveys is limited. The LAMOST survey observes the Northern Hemisphere, whereas RAVE and GALAH are focussed on the Southern Hemisphere, so in addition to being dedicated to hot stars, its intersection with the other ground surveys is also limited. The APOGEE footprint occupies both hemispheres.

It should be emphasised that a fraction of these comparisons can be contaminated by wrong cross-identifications when the different catalogues have been cross-matched \citep{2020ASPC..522..125P} by positions in the sky. Rotational broadening determinations can also be biased by non-detected spectroscopic companions, or by stellar activity, and these biases can affect the comparison catalogues differently depending on spectral coverage, resolving power, etc. 

\begin{figure*}[!tp]
{\center
\includegraphics[width=0.99\linewidth,clip=,draft=False,angle=0]{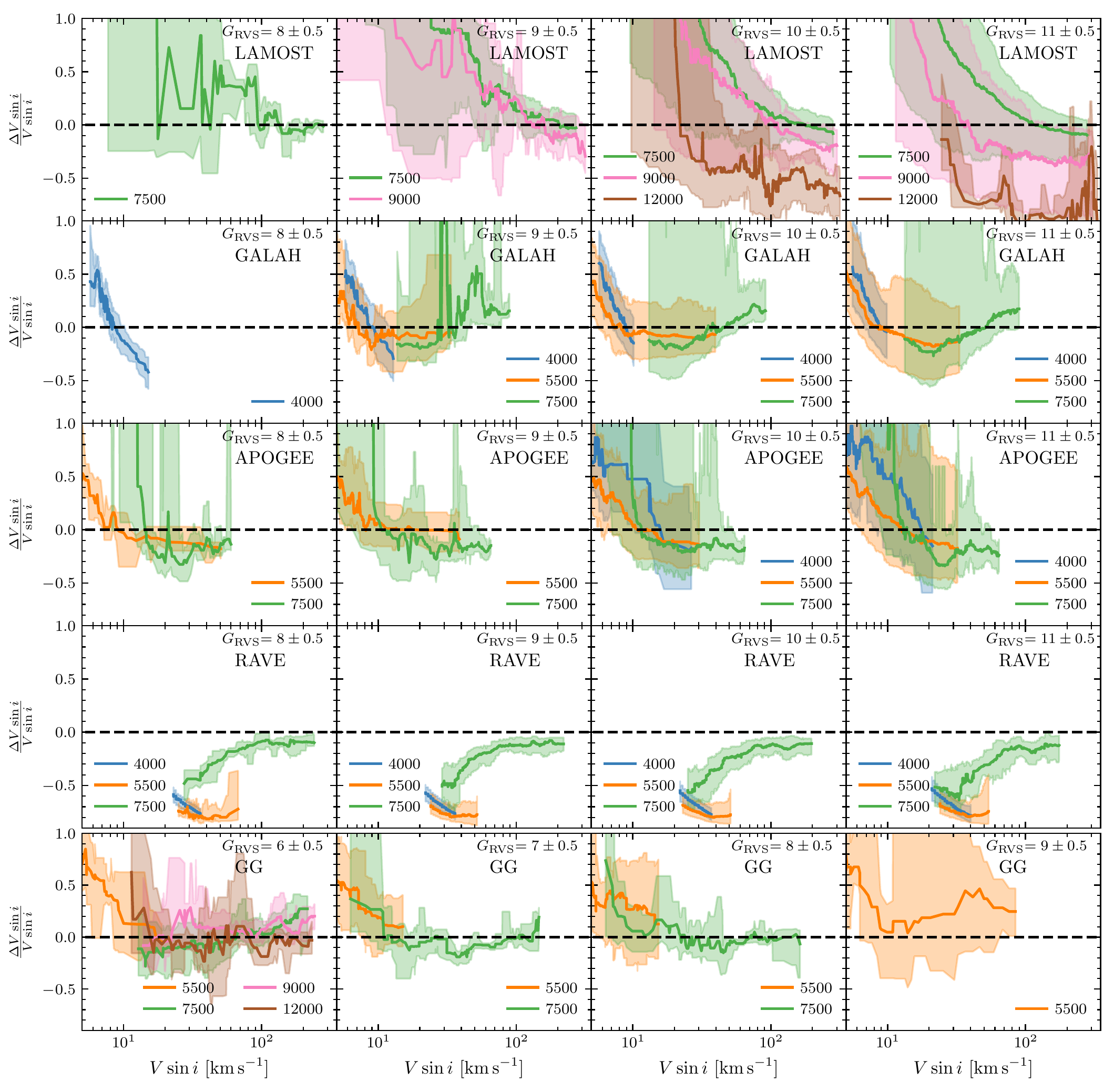}
\caption{Variation of the relative residuals in \vbroad\ as a function of the catalogue \vsini\ ($\Delta\vsini=\vbroad-\vsini$), for different ranges of effective temperature. The x-axis \vsini\ scales are the ones from the comparison catalogues. From left to right, panels inspect fainter ranges of magnitudes, respectively  7.5--8.5,  8.5--9.5 and 9.5--10.5\,mag, except for GG (last row) where magnitude ranges are shifted 2 mag brighter. Thick lines represent the running median on the residuals, while the coloured regions correspond to the associated 15\% and 85\% quantiles. Each colour corresponds to the temperature given in the plots.}\label{fig:residcat}
}
\end{figure*}

The logarithmic scales in Fig.\,\ref{fig:compvsini} allow us to overview the shifts at low and high line broadening values. Overdensities are present in the low velocity lower-left panel-corner for the comparison samples (GALAH, APOGEE and RAVE) dominated by cool slowly rotating stars. Comparisons with GALAH and APOGEE data, performed with a higher resolving power, show that the \vbroad\ determinations in \GDRthree\ are overestimated at lower \vsini\ partly due to the lower resolution in the RVS spectra. The spectral resolution in the RAVE survey is lower than in the RVS, and their rotational velocity determinations, in addition to being rounded to integer values, reach a plateau around 20\,\kps\ (only 2\% of the \vsini\ in the RAVE comparison sample are lower than 20\,\kps).

The right upper part of the panels is only populated with the catalogues that contain fast rotating stars and are able to determine high rotational velocities. The APOGEE survey has a hard upper limit at 96\,\kps, partly explaining why the lower-right quadrant of the \GDRthree-APOGEE panel is significantly populated. The comparison with LAMOST data is very dispersed, due to the much lower resolution and possibly the larger effect of template mismatch, but the catalogue content being mainly OBA targets ($\teff>7500$\,K) it allows an assessment of the \vbroad\ quality in the higher velocity range ($\vsini\gtrsim100$\,\kps). \elena{Whereas the comparison with GG seems in good agreement as soon as $\vbroad\gtrsim15$\,\kps, a trend appears at high values ($\vbroad\gtrsim200$\,\kps) where \vbroad\ determinations are systematically higher than their GG counterparts.}

The correlation and correspondences with the  catalogues we have considered tend to confirm that the \GDRthree\ \vbroad\ is a sensible measurement of the RVS line broadening. However, it also shares the  limitations at lower \vsini\ seen in other catalogues.

\subsection{Residuals as a function of \teff}

In order to quantify the residuals as a function of the observed magnitude and effective temperature, the comparison samples have been subsampled based on \grvs\ (\linktoparam{gaia_source}{grvs_mag}) and \teff\ (\linktoparam{gaia_source}{rv_template_teff}). It therefore gives a more detailed view of the trends visible in the first column of Fig.\,\ref{fig:compvsini}. The magnitude ranges are centred on $\grvs= 8$, 9, 10, and 11 (except those for GG, shifted 2 mag brighter), and have a width of 1 mag, while the effective temperature domains are taken at $\teff=4000\pm250$\,K, $5500\pm250$\,K, $7500\pm500$\,K, $9000\pm500$\,K, and $12\,000\pm1000$\,K.


Figure\,\ref{fig:residcat} shows the resulting distribution of the residuals with magnitude and effective temperature. We only plotted those subsamples with more than 80 targets, while the width of the running window represents one twelfth of the total number of measurements in the subsample. Only few comparison ensembles are able to provide information on the residuals for the coolest (\teff\ at 4000\, K) or the hottest targets (\teff\ range at 12\,000\,K). 

On average, when \teff\ subsamples are present at different magnitudes for the same catalogue, there is no significant impact on the residuals' offset, while their dispersion tends to increase with \grvs. 
As a global tendency, the residuals show that the \GDRthree\ \vbroad\ determinations are overestimated at low \vsini\ compared to other catalogues. By comparison with GG, GALAH, and APOGEE, this overestimation appears below $\sim12$\,\kps. At larger values, and if we exclude GG, \vbroad\ appears to underestimate \vsini\ by magnitudes that depend on \teff\ and \grvs.

Comparison with GG shows a good agreement for bright targets (6--8 mag), without any notable bias for velocities larger than $\sim$15\,\kps. At magnitude $\grvs=9$, GG is no more dominated by its largest contributors and starts being a compilation of only small heterogeneous datasets: the 127 targets that populate the right panel for GG in Fig.\,\ref{fig:residcat} may not be representative of the residual distribution. Moreover, the same \teff\ subsample at magnitude $\grvs\sim 9$ shows better agreement in comparisons with homogeneous catalogues such as GALAH or APOGEE.

Concerning the comparison with RAVE data, Fig.\,\ref{fig:compvsini} already showed that their low \vsini\ are systematically overestimated, whatever catalogue they are compared with. For velocities larger than $\sim$60\,\kps\ however, the residuals with \GDRthree\ \vbroad\ improve, being around $-10$\%, with a very small dispersion. This low scatter may result from the identical spectral range and similar resolving power as for RVS spectra.

The much lower resolving power in LAMOST spectra dominates the observed residuals below $\vsini\lesssim100$\,\kps. Above this value, the rotational broadening determinations are consistent for the \teff\ range 7500\,K, but the residuals significantly increase with magnitude for hotter targets.


\section{Discussion \label{sec:discussion}}

\begin{figure}[!htp]
\includegraphics[width=\linewidth,clip=,draft=False]{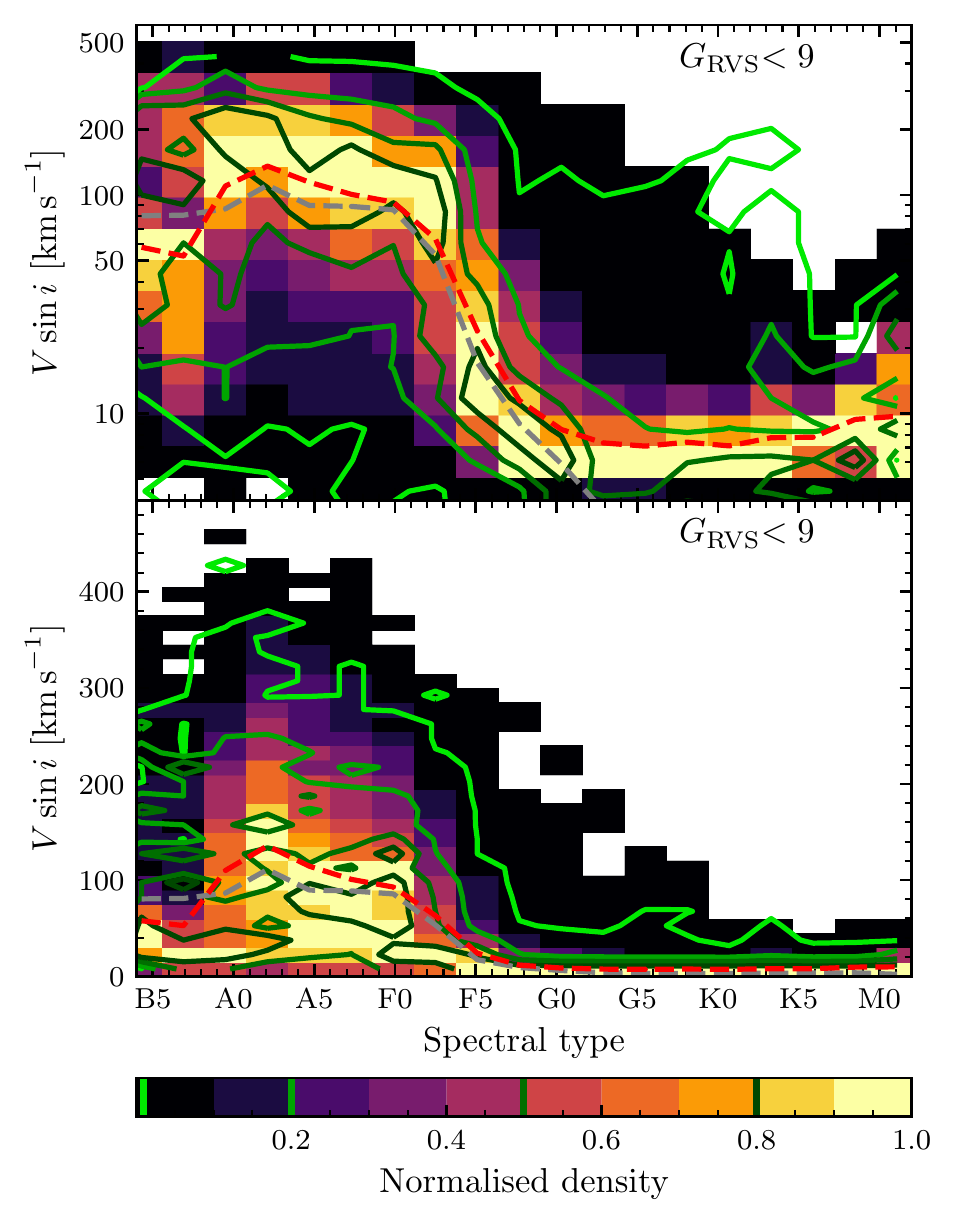}
\caption{Comparison of the distribution of \vbroad\ as a function of spectral type (2D histogram, coloured by the linear number of targets), with the distribution of \vsini\ from GG (green contour lines). Top panel shows the distribution with regular bins in logarithmic velocity scale, whereas bottom panel displays the resulting distribution using a linear grid in velocity.
The \vbroad\ data are selected to be brighter than $\grvs=9$ and to be on the main-sequence ($3.5\leq\logg\leq4.5$). \vsini\ data from GG are selected in the GG comparison sample (Table\,\ref{tab:catalogues}) with the same \logg\ criterion. Spectral types are estimated on the basis of \linktoparam{gaia_source}{rv_template_teff}, by interpolating in the tables provided by \citet{Allen}. Dashed lines are the median values per bin of spectral types, for the \vbroad\ distribution (red) and the \vsini\ (grey). For each bin of spectral type, the distribution is normalised to its maximum value. The colour bar superimposes the scale of the 2D histogram with the contour levels (0.01, 0.2, 0.5 and 0.8).}
\label{fig:vsini-sptype}
\end{figure}

\elena{Figure\,\ref{fig:vsini-sptype} displays the variation of the \vbroad\ distribution as a function of the spectral type, as already shown by \citet{2014psce.conf..256R}, and compares it with \vsini\ data from  the GG comparison sample. The coloured density plot is based on 63\,248 \vbroad\ values of main-sequence stars ($3.5\leq\logg\leq4.5$) brighter than $\grvs=9$. The contour plot is derived from 9262 \vsini\ values compiled by GG, with the same selection criterion on \logg.}

The modes of the distribution seem consistent between \vbroad\ and \vsini. \corr{The top panel in logarithmic scale reproduces the overestimation of \vbroad\ at low \vsini, already illustrated by Figs.\,\ref{fig:compvsini} and \ref{fig:residcat}, materialised here by  spectral types later than F5. In the bottom panel,} for hot stars, the contour low levels do not perfectly coincide with the \vbroad\ distribution counts, suggesting that high velocity distribution tails are more extended in the \GDRthree\ catalogue. As a result, the median values are also larger by 8 to 28\% from F0- to A0-type stars. This broadening of the \GDRthree\ data is produced by the trend observed between both velocity scales in Fig.\,\ref{fig:compvsini}. 


\begin{figure}[!htp]
\includegraphics[width=\linewidth,clip=,draft=False]{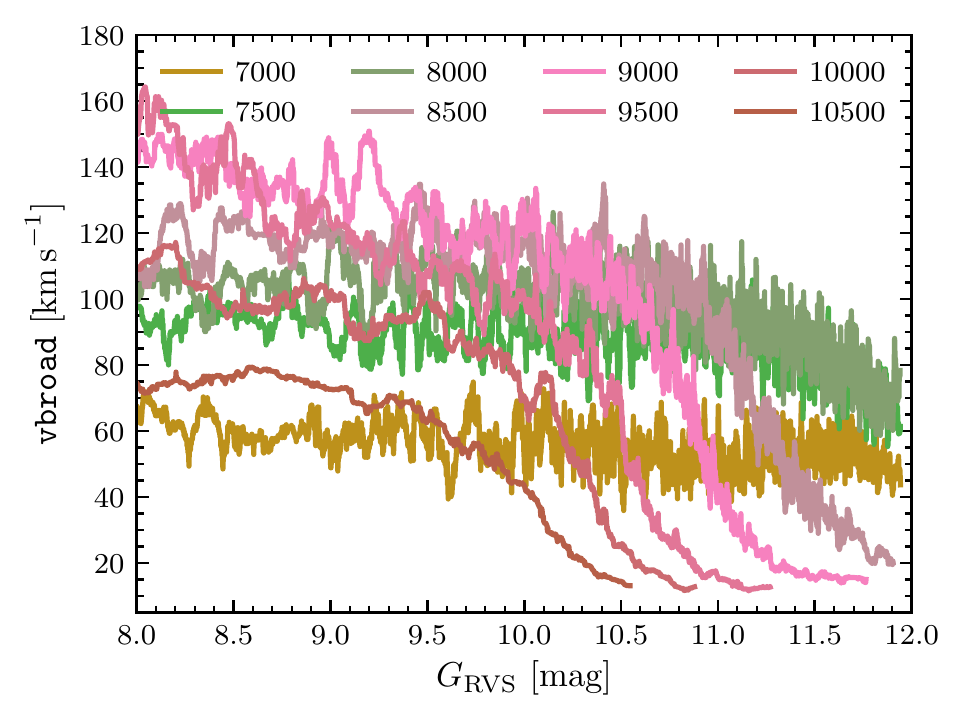}
\caption{Median value of \vbroad\ as a function \grvs\ for different \teff. Temperatures are taken as exact \linktoparam{gaia_source}{rv_template_teff} values and median \vbroad\ are derived on a running window of 200 points. Each colour corresponds to a \teff\ labelled in the plot.}
\label{fig:magnitudeeffect_hot}
\end{figure}


The catalogue-to-catalogue correlation and residual plots of Sect.\,\ref{sec:catalogues} reproduce the two main features identified during the MC simulations (Sect.\,\ref{sec:method:significance}). The \GDRthree\ \vbroad\ overestimates the low \vsini\ values, while it tends to underestimate the larger ones. From the simulations (Fig.\,\ref{fig:template_mismatch}), we noted for the hot stars a significant impact of the template mismatches due to a wrong \teff\ estimate. \corr{Still}, the comparisons made with the OBA LAMOST catalogue present, above 100 \kps, relative residuals (lower panels of Fig.\,\ref{fig:residcat}) which in magnitude are fairly consistent with those found in the simulations (Fig.\,\ref{fig:mc.rel.error.ms.sol}) when the effects of template mismatches are neglected. 

However, the simulations (e.g. Fig.\,\ref{fig:mc.abs.error.ms.sol}) also show that the quality of the results obtained above 7500\,K rapidly degrades with magnitude above $\grvs=10$.
\elena{In order to further investigate this degradation of the \vbroad\ quality with magnitude for hot stars, the median \vbroad\ is plotted as a function of \grvs\ for different \teff\ (Fig.\,\ref{fig:magnitudeeffect_hot}), exploring the transition from \ion{Ca}{ii} triplet dominated spectra to Paschen series dominated ones. There is no noticeable trend for 7000\,K stars (dark gold colour), whereas a slight decrease of \vbroad\ appears for 7500--8000\,K stars (greenish curves) at $\grvs\gtrsim 11$. For hotter stars (shades of pink), the effect is striking and increases with temperature. In addition to this severe underestimation of \vbroad\ at faint magnitude, we notice an apparent cut in \grvs, also increasing with temperature. This incompleteness was already seen in Fig.\,\ref{fig:teff.mag.dist} and is the combined result of the degradation of \vbroad\ at faint magnitude with the post-processing filtering that discarded values with $\vbroad<5$\,\kps.}

\elena{Because these findings are consistent with the trends noticed in Sect.\,\ref{sec:method:significance}, we used the MC simulation results to define a validity domain of the line broadening estimate and based on the quantities provided in the catalogue (i.e. {\tt rv\_template\_teff}, {\tt vbroad}, and {\tt vbroad\_error}). We give in Table\,\ref{tab:validity.MC}, the \vbroad\ domain where the measurement has more than 90\% of chances to be within 2-$\sigma$ (where $\sigma$ is assumed to be equal to the standard deviation) of \vsini. We provide these `validity' ranges as a function of \grvs\ and \teff. They represent the domains where the \vbroad\ measurement and its provided uncertainty are expected to be consistent with \vsini\ when template mismatches can be ignored.}

\begin{table}
\begin{center}
\caption{\vbroad\ validity domains derived from the MC simulations. The validity domain was defined as being the range of \vbroad\ values where the measurement has more than 90\% of chances to be within 2-$\sigma$ of \vsini. Limit values above 30 \kps\ were rounded to the nearest ten.}\label{tab:validity.MC}
\begin{tabular}{rrr@{\,}c@{\,}r|rrr@{\hskip 1mm}c@{\hskip 1mm}r}
\hline
\hline
 &  & \multicolumn{3}{c|}{\vbroad}  &  && \multicolumn{3}{c}{\vbroad} \\
\teff\ & \grvs & \multicolumn{3}{c|}{validity} & \teff\ & \grvs & \multicolumn{3}{c}{validity} \\
 $[\mathrm{K}]$ & & \multicolumn{3}{c|}{[\kps]} &  $[\mathrm{K}]$ & & \multicolumn{3}{c}{[\kps]}\\
\hline
\multirow{4}{*}{4000} & 8 & &~~~~~$>$& 18 & \multirow{4}{*}{7500} & 8 & 12 &--& 110 \\
&  9 & &~~~~~$>$& 14 & &  9 & 10 &--& 140 \\
& 10 & &~~~~~$>$& 12 & & 10 &  9 &--& 280 \\
& 11 & &~~~~~$>$& 10 & & 11 &  8 &--& 330 \\
\hline
\multirow{4}{*}{5500} & 8 & &~~~~~$>$& 18 &\multirow{4}{*}{9000} & 8 & 9 &--& 130 \\
&  9 & &~~~~~$>$& 14 & &  9 &  8 &--& 130 \\
& 10 & &~~~~~$>$& 11 & & 10 &  6 &--& 140 \\
& 11 & &~~~~~$>$&  9 & & 11 & 22 &--& 400 \\
\hline
&&&&&\multirow{4}{*}{12000} & 8 & 10 &--& 250 \\
&&&&& & 9 & 9 &--& 110 \\
&&&&& & 10 & 9 &--& 40 \\
&&&&& & 11 & 280 &--& 420 \\
\cline{6-10}
\end{tabular}
\end{center}
\end{table}

\begin{figure*}[!htb]
\includegraphics[width=1\linewidth,clip=,draft=False]{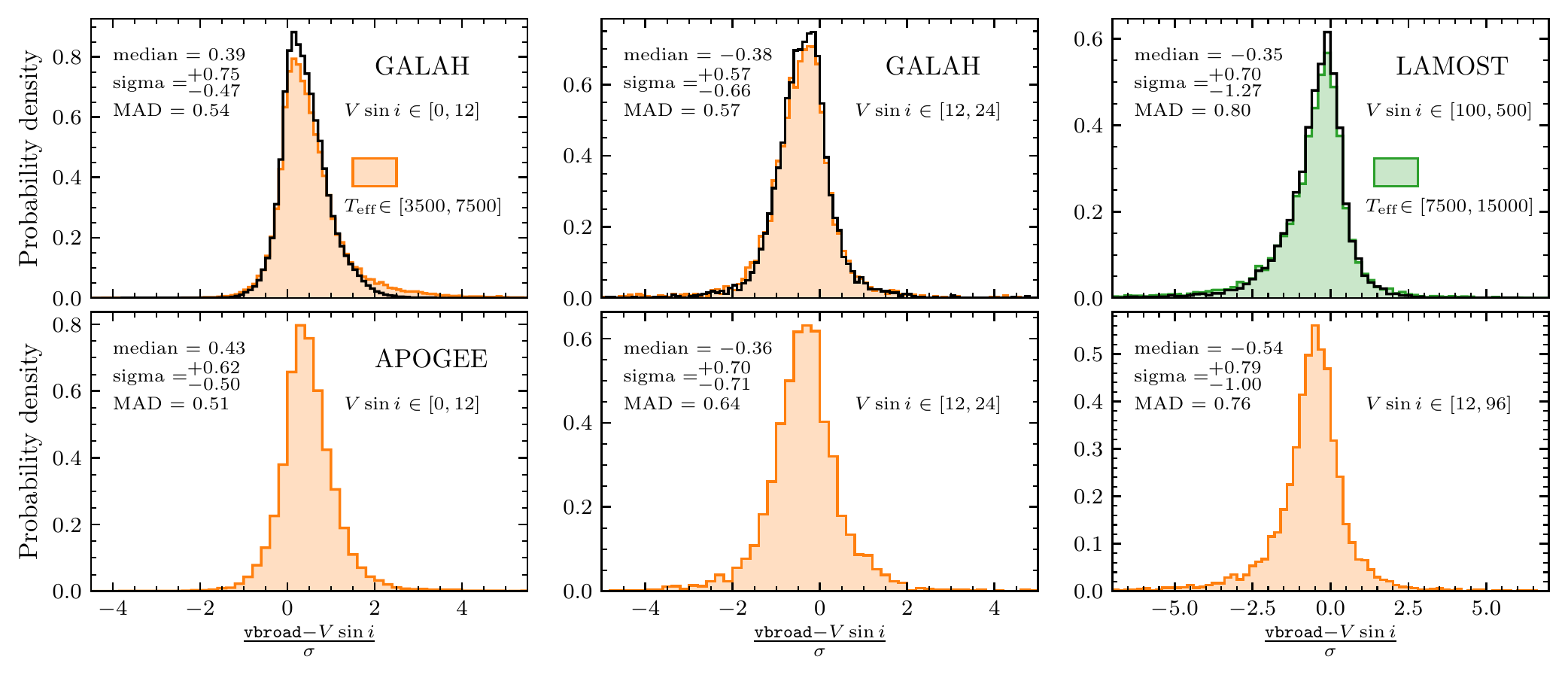}
 \caption{Distribution of the residuals for different catalogues: GALAH and LAMOST (top row) and APOGEE (bottom row). Residuals are normalised by the uncertainty on \vbroad\ in the \GDRthree\ catalogue. For the top panels, the superimposed black curve is the residual distribution normalised by the total uncertainty $\sqrt{\sigma_\vbroad^2 + \sigma_{\vsini}^2}$. Each row corresponds to a selection in \teff\ and \vsini. Statistical estimators are given for each panel: median value, upper and lower dispersions (85\% quantile$ - $median, and median$ - $15\% quantile) and mean absolute deviation. \label{fig:zscore}}
\end{figure*}

The \vbroad\ published in \GDRthree\ is the median value of a sample of $N_\mathrm{t}$ measurements (where the median value of the number of transits is 12 as shown in Fig.\,\ref{fig:nt_cdf}) done on transit spectra. During the validation, we decided to adopt their standard deviation as a measure of the uncertainty (Fig.\,\ref{fig:relsigvsGrvs}). In Fig.\,\ref{fig:zscore}, we confront this uncertainty to the scatter of the residuals of \vbroad\ to the \vsini\ measurements published in those catalogues (\elena{GALAH} and APOGEE) or \vsini\ ranges (LAMOST) which are expected to be less impacted by resolving power issues. We considered two \teff\ domains representative of the spectroscopic content of the RVS, as well as various \vsini\ domains. On the basis of the dispersions measured in the residual distributions, we note that the uncertainty provided for the F-, G-, and K-type stars in the catalogue can be overestimated by a factor of $\sim$2 in the low \vbroad\ regime and by a factor of $\sim$1.3 for larger \vbroad\ estimates. 
On the other hand, for the hotter stars, the uncertainty tends to be less overestimated (i.e. by a factor of $\sim$1.25). 

Both GALAH and LAMOST catalogues provide uncertainty estimates for the derived \vsini, which offers the possibility to quantify the change in z-score as a function of magnitude. As Fig.\ref{fig:zscore} shows residual distributions representative of the full common magnitude range with the catalogue, Table\,\ref{tab:zscores_galah} lists the z-score results for the same \vsini\ ranges and different magnitude intervals. For the cool stars, the dispersion decreases from $\sim$0.9 to $\sim$0.5 as magnitudes get faint. It suggests that uncertainty in the \GDRthree\ \vbroad\ values are even more overestimated as we go faint. 
For the hotter, fast rotating stars, the comparison with LAMOST indicates that \vbroad\ uncertainty in the \GDRthree\ catalogue is probably underestimated for stars brighter than $\grvs=10$, but overestimated for fainter stars. It is worth keeping in mind that the LAMOST comparison sample is dominated by stars with \teff\ around 8000\,K (Fig.\,\ref{fig:cats.overview}), and the effect illustrated in Fig.\ref{fig:magnitudeeffect_hot} solely contributes to the tails of the z-score distribution. 
Figure\,\ref{fig:relsigvsGrvs} displays the average relative uncertainties at magnitudes $\grvs=8,9,10$ and 11, taking into account the MAD values from Table\,\ref{tab:zscores_galah} as correction factors. 

\begin{table}[!tp]
    \centering
    \caption{z-score statistics from the comparison with the GALAH and LAMOST catalogues, normalised by the total uncertainty, for different ranges of magnitude and different ranges of \vsini. The median value of $(\vbroad-\vsini)/\sigma$, with $\sigma=\sqrt{\sigma_\vbroad^2 + \sigma_{\vsini}^2}$,  is given as $z_\mathrm{med}$; $\sigma_{z+}$  and $\sigma_{z-}$ are the upper and lower dispersions (85\% quantile$ -z_\mathrm{med}$, and 15\% quantile$ -z_\mathrm{med}$); MAD is the mean absolute deviation, and \# the number of targets in the corresponding subsample.}
    \label{tab:zscores_galah}
    \begin{tabular}{r@{\,}c@{\,}rrrrrr}
    \hline\hline
   \multicolumn{3}{c}{\grvs}   &  $z_\mathrm{med}$  &  $\sigma_{z+}$  &  $\sigma_{z-}$  &   MAD   &     \multicolumn{1}{c}{\#}  \\
   \hline
\multicolumn{8}{c}{GALAH: $\vsini\in [ 0, 12]$\,\kps}\\
\hline
  $7.5$&--&$8.5$  &    0.80     &    0.85   &       $-0.84$ &    0.86  &     784\\
  $8.5$&--&$9.5$  &    0.69     &    0.84   &       $-0.78$ &    0.80  &    3884\\
 $9.5$&--&$10.5$  &    0.50     &    0.75   &       $-0.63$ &    0.65  &   13306\\
$10.5$&--&$11.5$  &    0.33     &    0.52   &       $-0.42$ &    0.45  &   33760\\
   \hline
\multicolumn{8}{c}{GALAH: $\vsini\in [12,24]$\,\kps}\\
\hline
  $7.5$&--&$8.5$  &   $-1.99$   &      1.02 &         $-1.38$&     0.96 &       48\\
  $8.5$&--&$9.5$  &   $-0.86$   &      0.75 &         $-1.20$&     0.82 &      238\\
 $9.5$&--&$10.5$  &   $-0.59$   &      0.75 &         $-0.61$&     0.60 &     1125\\
$10.5$&--&$11.5$  &   $-0.35$   &      0.53 &         $-0.55$&     0.50 &     3006\\
\hline
\multicolumn{8}{c}{LAMOST: $\vsini\in [100,500]$\,\kps}\\
\hline
  $7.5$&--&$8.5$  &    $-0.34$    &     1.98 &         $-1.73$&     1.41 &       53\\
  $8.5$&--&$9.5$  &    $-0.18$    &     1.55 &         $-1.57$&     1.49 &      363\\
 $9.5$&--&$10.5$  &    $-0.34$    &     0.93 &         $-1.29$&     1.00 &     2198\\
$10.5$&--&$11.5$  &    $-0.31$    &     0.57 &         $-0.98$&     0.67 &     5545\\

\hline

    \end{tabular}
\end{table}


\begin{figure*}[!htb]
\includegraphics[width=\textwidth]{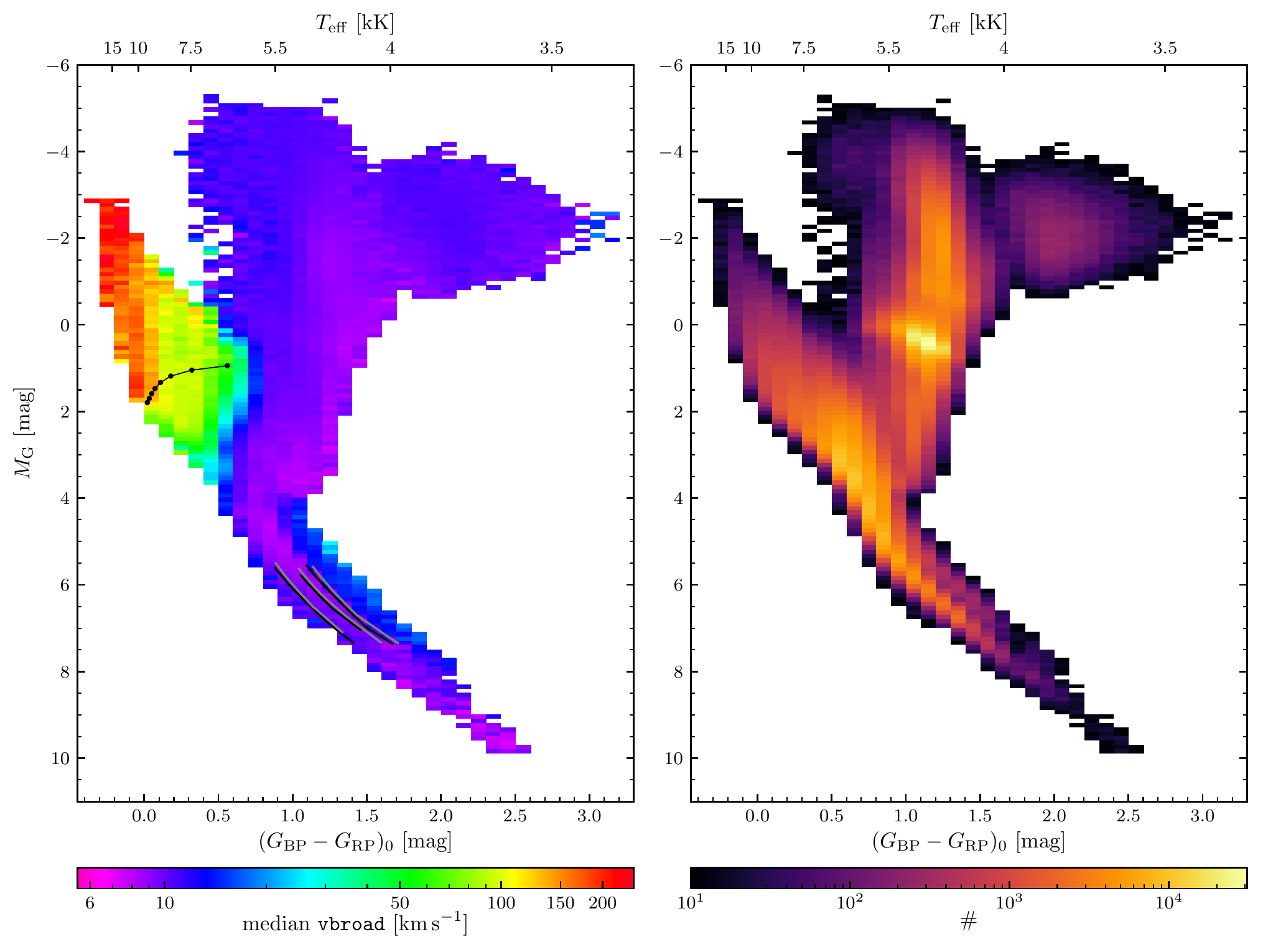}
\caption{Hertzsprung-Russell diagrams for a subsample of the \GDRthree\ \vbroad\ catalogue ($\sim 1.8$ million stars). The larger part of missing data is due to the lack of extinction parameters to correct $M_\mathrm{G}$ and deredden \bprp, for about 43\% of the sample. An additional cut is performed on the parallax quality ($\varpi/\sigma_\varpi>10$) and removes 3.2\% of the total sample. For hot stars, a selection is done on \grvs, removing an additional 2.5\% of the sample (see text).  The binning size is 0.1 by 0.1\,mag. Bins containing less than ten stars are discarded. Left panel maps the median \vbroad\ values (in logarithmic colour scale), whereas right panel shows the density, in order to better associate the rotational velocity map to the corresponding structures in the HRD. To guide the eye, the upper x-axes show the approximate \teff\ scale, calibrated as a function of \bprp\ using the photometric temperatures. Also, in the left panel, an evolutionary track of a $2\,\mathrm{M}_\sun$ star, sampled each 162.5\,Myr, illustrates the course from ZAMS to TAMS in the upper MS. In addition, three pairs of isochrones are superimposed to the lower main sequence, for two different ages (1\,Gyr in black and 10\,Gyr in grey), and three different metallicities: $\mathrm{[M/H]}=-0.5, 0, +0.5$, from left to right.}
\label{fig:HRD}
\end{figure*}

\corr{The final step of the validation shows the mapping of the median \vbroad\ across the Hertzsprung-Russell diagram (HRD, see Fig.\,\ref{fig:HRD}), using integrated photometry in the
$G$, $G_\mathrm{BP}$ and $G_\mathrm{RP}$ bands \citep{EDR3-DPACP-117}. For more than half the sample, extinction parameters are available from the {\it Apsis} pipeline \citep{DR3-DPACP-157,DR3-DPACP-160,DR3-DPACP-156}. The absorption in the $G$ band, $A_\mathrm{G}$, and the \bprp\ colour excess, $E(\bprp)$, are taken from \textit{ESP-HS} 
\citep{DR3-DPACP-157} for hot stars ($\teff>7500$\,K, \linktoaps{ag_esphs}, \linktoaps{ebpminrp_esphs}) and from \textit{GSP-Phot} for cooler ones (\linktoparam{gaia_source}{ag_gspphot}, \linktoparam{gaia_source}{ebpminrp_gspphot}). Both $A_\mathrm{G}$ and $E(\bprp)$ are taken into account to derive the positions $(\bprp)_0$ and $M_\mathrm{G}$ in the HRD. Only stars with parallaxes having a precision better than 10\% are shown in Fig.\,\ref{fig:HRD}. 
To limit the bias on \vbroad\ observed for hot stars in Fig.\,\ref{fig:magnitudeeffect_hot}, a filter in \grvs\ depending on \teff\ only, has been preferred to using the validity domains listed in Table\,\ref{tab:validity.MC}. Those domains would have biased the statistical values in the HRD. The applied filtering limit varies linearly as a function of \teff:
\begin{equation}
   \grvs<11.93 -0.8087\,10^{-3}\,(\teff-7000),\mbox{~for  $\teff>7000$\,K.}
    \label{eqn:grvscut_hot}
\end{equation}
The $0.1\times0.1$\,mag bins in the HRD are plotted only if they contain at least 10 stars. 
The diagram illustrates the large coverage of the parameter space by the \GDRthree\ \vbroad\ catalogue: evolutionary stages from the MS to the giant branch and the supergiants are present. The temperature scale in Fig.\,\ref{fig:HRD} is given as an indication, and it is based on the photometric temperatures, selected with the same criterion as for the extinction parameters (\linktoparam{gaia_source}{teff_gspphot} for $\teff\le7500$\,K, and
\linktoaps{teff_esphs} for $\teff>7500$\,K). It roughly corresponds to the \teff\ range 3500--14\,500\,K resulting from the applied filters (Sect.\,\ref{sec:method:filters}) on \linktoparam{gaia_source}{rv_template_teff}. \\
The most prominent feature in the left panel is due to the rapid drop of the mean rotational velocity of stars around spectral type F5, known since \citet{Kraft1967}, and already seen in Fig.\,\ref{fig:vsini-sptype} for MS stars. More massive stars are generally rapid rotators, while less massive ones are characterised by a slow rotation.
The evolutionary track of a solar metallicity $2\,\mathrm{M}_\sun$ star, generated by CMD 3.6\footnote{\url{http://stev.oapd.inaf.it/cmd}} \citep{Bressan2012,Chen2014,Chen2015,Tang2014,Marigo2017,Pastorelli2019,Pastorelli2020}, is overplotted to the upper MS from ZAMS (Zero Age Main Sequence) to TAMS (Terminal Age Main Sequence), as a reference.   \\
The lower MS in the right panel ($M_\mathrm{G}>5$) reveals the presence of the binary star sequence, $0.75$\,mag brighter than the MS of single stars. This sequence displays higher \vbroad\ values in the left panel. In the range $1.1<\bprp<1.4$\, for example, the median \vbroad\ values for the single sequence and the binary sequence are respectively 9 and 14\,\kps.
\\
The lower MS of single stars seems to harbour a decrease of velocity from left to right. The overplotted isochrones, generated by CMD 3.6, correspond to two different ages (1\,Gyr in black, 10\,Gyr in grey) and three different metallicities: $[M/H] = -0.5, 0, +0.5$, from left to right. They illustrate the fact that the thickness of the lower MS is dominated by a spread in the metallicity distribution rather than an evolutionary effect. It suggests that this trend in \vbroad\ could be due to mismatches in metallicity between the spectra and the templates: a template broadened with a lower \vbroad\ value has deeper lines and can better fit an observed spectrum with a higher metallicity. Hence it rules out the possibility of using the \Gaia\ DR3 \vbroad\ values as a gyrochronological tool and of inferring anything about stellar ages.
}

\subsection{Effect of the macroturbulence}

\corr{When measuring \vbroad, no distinction is made between stellar rotation and other mechanisms that contribute to broadening the spectral lines at constant equivalent width. In particular, no effort is done to remove, nor derive, the contribution of the {macroturbulence}. However, \vmacro\ is expected to have varying magnitude throughout the HRD. In late-type stars its origin and impact is explained by surface convection and by 3D modelling \citep{2000A&A...359..729A}. At hotter temperatures, observations suggest that the origin of \vmacro\ might be various competing phenomena: it can be related to line-profile variations \citep{2009A&A...508..409A} due to surface inhomogeneity and pulsation, or to turbulent pressure \citep{2015ApJ...808L..31G}. {Macroturbulence} is usually expected to broaden the line shapes with a Gaussian-like kernel, and requires high S/N and high spectral resolution data to be accurately disentangled from the rotational broadening. Obviously, these conditions are not met by the epoch \GDRthree\ RVS data. Accurate measurements based on 1D stellar atmosphere modelling show that its value increases with temperature and luminosity. In the \teff\ range covered by the \GDRthree\ \vbroad\ catalogue (i.e. $3500 \le \teff\ \le 14\,500$\,K), the macroturbulence is thought to increase with \teff, and with decreasing \logg\ \citep{2014MNRAS.444.3592D}. It has values of the order of 2 to 3.5\,\kps\ at 5000\,K, and 5 to 6.5\,\kps\ at 6400\,K. At the hottest edge, B-type supergiants usually have their line-broadening dominated by \vmacro\ with values larger than 25\,\kps\ \citep{2017A&A...597A..22S}. At lower luminosity, \vmacro\ tends to be lower than \vsini\ but can still have values as large as $\sim$60\,\kps.}

\subsection{Effect of ignored binarity}

\corr{
A spectroscopically unresolved companion can also impact the measurements. According to \citet{2014ApJ...788L..37G}, and based on a sample of binaries with periods less than 1000\,days, the overall fraction of FGK binary systems in the Milky Way has a value expected to lie in the range of 0.30 to 0.56, depending on metallicity and on the data that were adopted to infer it. In Solar-type stars and for close binaries \citep{2019ApJ...875...61M}, it was found to be anti-correlated with metallicity varying from 0.53 to 0.24 for $\mathrm{[Fe/H]} = -3$ to $-0.2$, respectively. Furthermore, this fraction of multiple systems is known to increase with mass and is observed to reach a value of 0.91 to 1 in O-type stars \citep[][reminding here that the \GDRthree\ \vbroad\ catalogue does not include stars earlier than 14\,500\,K]{Sana_2014}. During the processing and analysis of the RVS spectra a significant effort was made to flag the double-lined spectroscopic binaries \citep{DR3-DPACP-161,DR3-DPACP-159}, and to remove from the catalogue their median RV and \vbroad\ estimates. As shown in Fig.\,\ref{fig:HRD}, some binaries survived the post-processing cleaning. A counting of the sources in part of the single and binary star main sequences (\bprp\ ranging from 1.1 to 1.4) 
provides a fraction of 0.17 of MS `candidate' multiple stars that would still have a published \vbroad\ estimate. A random visual inspection of the corresponding RVS spectra shows that, while known spectroscopic binaries are found in the sample, most of these were not spectroscopically resolved. We may expect, from such `hidden' binarity, line profile and strength variability (e.g. panel f in Fig.\,\ref{fig:rvsspectra}) that produces, statistically, a global overestimate of the line-broadening as Fig.\,\ref{fig:HRD} suggests.
}

\section{Conclusions\label{sec:conclusions}}

The \GDRthree\ catalogue provides the largest survey of line broadening estimates down to magnitude 12, and from 3500\,K to 14\,500\,K (Fig.\,\ref{fig:teff.mag.dist}). These estimates include all the line broadening terms that are not taken into account by the synthetic spectra (e.g. \vsini, macroturbulence). Therefore, as was done in other surveys (e.g. GALAH), we named the measurement \vbroad.

While our validation work globally shows that the measurements are fairly consistent with other surveys and compilations, it also reminds that the choice of the RVS wavelength domain was optimised to allow the RV measurement of most Gaia targets, but not for their accurate and non-biased determination of \vsini. This is especially the case for the stars hotter than 7500\,K, when the features that dominate the spectrum are due to the intrinsically broad lines of the hydrogen Paschen series and of the \ion{Ca}{ii} triplet. By nature, these features are strongly blended and their relative dependence on the astrophysical parameters may lead to template mismatches to which the determination of \vbroad\ is quite sensitive. As confirmed by the catalogue-to-catalogue comparisons, their impact was mitigated by the use of updated APs obtained for the hot stars during the RV processing \citep{DR3-DPACP-151}. However, at $\teff>7500$\,K the dependence of the \vbroad\ accuracy and precision with temperature and \grvs\ remains complex and degrades rapidly above $\grvs=10$. 
\corr{The colour-magnitude diagram (Fig.\,\ref{fig:HRD}) shows how the median \vbroad\ varies in the HRD. While it reproduces the main feature expected due to magnetic braking in the cool stars around F5, it also highlights a potential effect of a mismatch due to metallicity between the observed spectrum and the template used to derive the value.}
Therefore, in general, we recommend to remain cautious in the interpretation of the \vbroad\ parameter values. To better help the catalogue user, we provide in Table\,\ref{tab:validity.MC} an estimate of the \vbroad\ domains where both \vbroad\ and its uncertainty are expected to be consistent with \vsini. 


During the processing of \GDRthree, the \vbroad\ results obtained by the method described in Sect.\,\ref{sec:method:description} were considered. More tests will be conducted during the preparation of the next data release in order to include the estimates from other algorithms (e.g. minimum distance method, use of the CCF width, ...). The method presented in this paper uses the information integrated over the complete RVS domain (i.e. it produces one single CCF). It has obvious advantages for the fainter targets, but it is usually also dominated by the stronger and broader features which are less sensitive to any additional line broadening. Therefore, among the tests we conduct to prepare \Gaia\ DR4, we check the pertinence of isolating certain portions of the spectra more sensitive to the rotational broadening and of performing the measurement on co-added spectra.


\begin{acknowledgements}
We thank Dr Elena Pancino and the anonymous referee for carefully reading the manuscript and for providing us with  constructive comments that helped to improve the paper.

This work presents results from the European Space Agency (ESA) space mission \Gaia. \Gaia\ data are being processed by the \Gaia\ Data Processing and Analysis Consortium (DPAC). Funding for the DPAC is provided by national institutions, in particular the institutions participating in the \Gaia\ MultiLateral Agreement (MLA). The \Gaia\ mission website is \url{https://www.cosmos.esa.int/gaia}. The \Gaia\ archive website is \url{https://archives.esac.esa.int/gaia}.
Acknowledgements are given in Appendix\,\ref{ssec:appendixA}.

This work has used the following software products:
Matplotlib \citep[][\url{https://matplotlib.org}]{Hunter:2007},
SciPy \citep[][\url{https://www.scipy.org}]{2020SciPy-NMeth}, and
NumPy \citep[][\url{https://numpy.org}]{harris2020array}.  
This research made use of the SIMBAD database, the Vizier catalogue access, and the cross-match services provided by CDS, Strasbourg, France.
\end{acknowledgements}

%
%


\bibliographystyle{naa}
\bibliography{gaia,dpac,other}

\begin{appendix}
\section{}\label{ssec:appendixA}
This work presents results from the European Space Agency (ESA) space mission \Gaia. \Gaia\ data are being processed by the \Gaia\ Data Processing and Analysis Consortium (DPAC). Funding for the DPAC is provided by national institutions, in particular the institutions participating in the \Gaia\ MultiLateral Agreement (MLA). The \Gaia\ mission website is \url{https://www.cosmos.esa.int/gaia}. The \Gaia\ archive website is \url{https://archives.esac.esa.int/gaia}.

The \Gaia\ mission and data processing have financially been supported by, in alphabetical order by country:
\begin{itemize}
\item the Algerian Centre de Recherche en Astronomie, Astrophysique et G\'{e}ophysique of Bouzareah Observatory;
\item the Austrian Fonds zur F\"{o}rderung der wissenschaftlichen Forschung (FWF) Hertha Firnberg Programme through grants T359, P20046, and P23737;
\item the BELgian federal Science Policy Office (BELSPO) through various PROgramme de D\'{e}veloppement d'Exp\'{e}riences scientifiques (PRODEX) grants and the Polish Academy of Sciences - Fonds Wetenschappelijk Onderzoek through grant VS.091.16N, and the Fonds de la Recherche Scientifique (FNRS), and the Research Council of Katholieke Universiteit (KU) Leuven through grant C16/18/005 (Pushing AsteRoseismology to the next level with TESS, GaiA, and the Sloan DIgital Sky SurvEy -- PARADISE);  
\item the Brazil-France exchange programmes Funda\c{c}\~{a}o de Amparo \`{a} Pesquisa do Estado de S\~{a}o Paulo (FAPESP) and Coordena\c{c}\~{a}o de Aperfeicoamento de Pessoal de N\'{\i}vel Superior (CAPES) - Comit\'{e} Fran\c{c}ais d'Evaluation de la Coop\'{e}ration Universitaire et Scientifique avec le Br\'{e}sil (COFECUB);
\item the Chilean Agencia Nacional de Investigaci\'{o}n y Desarrollo (ANID) through Fondo Nacional de Desarrollo Cient\'{\i}fico y Tecnol\'{o}gico (FONDECYT) Regular Project 1210992 (L.~Chemin);
\item the National Natural Science Foundation of China (NSFC) through grants 11573054, 11703065, and 12173069, the China Scholarship Council through grant 201806040200, and the Natural Science Foundation of Shanghai through grant 21ZR1474100;  
\item the Tenure Track Pilot Programme of the Croatian Science Foundation and the \'{E}cole Polytechnique F\'{e}d\'{e}rale de Lausanne and the project TTP-2018-07-1171 `Mining the Variable Sky', with the funds of the Croatian-Swiss Research Programme;
\item the Czech-Republic Ministry of Education, Youth, and Sports through grant LG 15010 and INTER-EXCELLENCE grant LTAUSA18093, and the Czech Space Office through ESA PECS contract 98058;
\item the Danish Ministry of Science;
\item the Estonian Ministry of Education and Research through grant IUT40-1;
\item the European Commission’s Sixth Framework Programme through the European Leadership in Space Astrometry (\href{https://www.cosmos.esa.int/web/gaia/elsa-rtn-programme}{ELSA}) Marie Curie Research Training Network (MRTN-CT-2006-033481), through Marie Curie project PIOF-GA-2009-255267 (Space AsteroSeismology \& RR Lyrae stars, SAS-RRL), and through a Marie Curie Transfer-of-Knowledge (ToK) fellowship (MTKD-CT-2004-014188); the European Commission's Seventh Framework Programme through grant FP7-606740 (FP7-SPACE-2013-1) for the \Gaia\ European Network for Improved data User Services (\href{https://gaia.ub.edu/twiki/do/view/GENIUS/}{GENIUS}) and through grant 264895 for the \Gaia\ Research for European Astronomy Training (\href{https://www.cosmos.esa.int/web/gaia/great-programme}{GREAT-ITN}) network;
\item the European Cooperation in Science and Technology (COST) through COST Action CA18104 `Revealing the Milky Way with \Gaia\ (MW-Gaia)';
\item the European Research Council (ERC) through grants 320360, 647208, and 834148 and through the European Union’s Horizon 2020 research and innovation and excellent science programmes through Marie Sk{\l}odowska-Curie grant 745617 (Our Galaxy at full HD -- Gal-HD) and 895174 (The build-up and fate of self-gravitating systems in the Universe) as well as grants 687378 (Small Bodies: Near and Far), 682115 (Using the Magellanic Clouds to Understand the Interaction of Galaxies), 695099 (A sub-percent distance scale from binaries and Cepheids -- CepBin), 716155 (Structured ACCREtion Disks -- SACCRED), 951549 (Sub-percent calibration of the extragalactic distance scale in the era of big surveys -- UniverScale), and 101004214 (Innovative Scientific Data Exploration and Exploitation Applications for Space Sciences -- EXPLORE);
\item the European Science Foundation (ESF), in the framework of the \Gaia\ Research for European Astronomy Training Research Network Programme (\href{https://www.cosmos.esa.int/web/gaia/great-programme}{GREAT-ESF});
\item the European Space Agency (ESA) in the framework of the \Gaia\ project, through the Plan for European Cooperating States (PECS) programme through contracts C98090 and 4000106398/12/NL/KML for Hungary, through contract 4000115263/15/NL/IB for Germany, and through PROgramme de D\'{e}veloppement d'Exp\'{e}riences scientifiques (PRODEX) grant 4000127986 for Slovenia;  
\item the Academy of Finland through grants 299543, 307157, 325805, 328654, 336546, and 345115 and the Magnus Ehrnrooth Foundation;
\item the French Centre National d’\'{E}tudes Spatiales (CNES), the Agence Nationale de la Recherche (ANR) through grant ANR-10-IDEX-0001-02 for the `Investissements d'avenir' programme, through grant ANR-15-CE31-0007 for project `Modelling the Milky Way in the \Gaia\ era’ (MOD4Gaia), through grant ANR-14-CE33-0014-01 for project `The Milky Way disc formation in the \Gaia\ era’ (ARCHEOGAL), through grant ANR-15-CE31-0012-01 for project `Unlocking the potential of Cepheids as primary distance calibrators’ (UnlockCepheids), through grant ANR-19-CE31-0017 for project `Secular evolution of galaxies' (SEGAL), and through grant ANR-18-CE31-0006 for project `Galactic Dark Matter' (GaDaMa), the Centre National de la Recherche Scientifique (CNRS) and its SNO \Gaia\ of the Institut des Sciences de l’Univers (INSU), its Programmes Nationaux: Cosmologie et Galaxies (PNCG), Gravitation R\'{e}f\'{e}rences Astronomie M\'{e}trologie (PNGRAM), Plan\'{e}tologie (PNP), Physique et Chimie du Milieu Interstellaire (PCMI), and Physique Stellaire (PNPS), the `Action F\'{e}d\'{e}ratrice \Gaia' of the Observatoire de Paris, the R\'{e}gion de Franche-Comt\'{e}, the Institut National Polytechnique (INP) and the Institut National de Physique nucl\'{e}aire et de Physique des Particules (IN2P3) co-funded by CNES;
\item the German Aerospace Agency (Deutsches Zentrum f\"{u}r Luft- und Raumfahrt e.V., DLR) through grants 50QG0501, 50QG0601, 50QG0602, 50QG0701, 50QG0901, 50QG1001, 50QG1101, 50\-QG1401, 50QG1402, 50QG1403, 50QG1404, 50QG1904, 50QG2101, 50QG2102, and 50QG2202, and the Centre for Information Services and High Performance Computing (ZIH) at the Technische Universit\"{a}t Dresden for generous allocations of computer time;
\item the Hungarian Academy of Sciences through the Lend\"{u}let Programme grants LP2014-17 and LP2018-7 and the Hungarian National Research, Development, and Innovation Office (NKFIH) through grant KKP-137523 (`SeismoLab');
\item the Science Foundation Ireland (SFI) through a Royal Society - SFI University Research Fellowship (M.~Fraser);
\item the Israel Ministry of Science and Technology through grant 3-18143 and the Tel Aviv University Center for Artificial Intelligence and Data Science (TAD) through a grant;
\item the Agenzia Spaziale Italiana (ASI) through contracts I/037/08/0, I/058/10/0, 2014-025-R.0, 2014-025-R.1.2015, and 2018-24-HH.0 to the Italian Istituto Nazionale di Astrofisica (INAF), contract 2014-049-R.0/1/2 to INAF for the Space Science Data Centre (SSDC, formerly known as the ASI Science Data Center, ASDC), contracts I/008/10/0, 2013/030/I.0, 2013-030-I.0.1-2015, and 2016-17-I.0 to the Aerospace Logistics Technology Engineering Company (ALTEC S.p.A.), INAF, and the Italian Ministry of Education, University, and Research (Ministero dell'Istruzione, dell'Universit\`{a} e della Ricerca) through the Premiale project `MIning The Cosmos Big Data and Innovative Italian Technology for Frontier Astrophysics and Cosmology' (MITiC);
\item the Netherlands Organisation for Scientific Research (NWO) through grant NWO-M-614.061.414, through a VICI grant (A.~Helmi), and through a Spinoza prize (A.~Helmi), and the Netherlands Research School for Astronomy (NOVA);
\item the Polish National Science Centre through HARMONIA grant 2018/30/M/ST9/00311 and DAINA grant 2017/27/L/ST9/03221 and the Ministry of Science and Higher Education (MNiSW) through grant DIR/WK/2018/12;
\item the Portuguese Funda\c{c}\~{a}o para a Ci\^{e}ncia e a Tecnologia (FCT) through national funds, grants SFRH/\-BD/128840/2017 and PTDC/FIS-AST/30389/2017, and work contract DL 57/2016/CP1364/CT0006, the Fundo Europeu de Desenvolvimento Regional (FEDER) through grant POCI-01-0145-FEDER-030389 and its Programa Operacional Competitividade e Internacionaliza\c{c}\~{a}o (COMPETE2020) through grants UIDB/04434/2020 and UIDP/04434/2020, and the Strategic Programme UIDB/\-00099/2020 for the Centro de Astrof\'{\i}sica e Gravita\c{c}\~{a}o (CENTRA);  
\item the Slovenian Research Agency through grant P1-0188;
\item the Spanish Ministry of Economy (MINECO/FEDER, UE), the Spanish Ministry of Science and Innovation (MICIN), the Spanish Ministry of Education, Culture, and Sports, and the Spanish Government through grants BES-2016-078499, BES-2017-083126, BES-C-2017-0085, ESP2016-80079-C2-1-R, ESP2016-80079-C2-2-R, FPU16/03827, PDC2021-121059-C22, RTI2018-095076-B-C22, and TIN2015-65316-P (`Computaci\'{o}n de Altas Prestaciones VII'), the Juan de la Cierva Incorporaci\'{o}n Programme (FJCI-2015-2671 and IJC2019-04862-I for F.~Anders), the Severo Ochoa Centre of Excellence Programme (SEV2015-0493), and MICIN/AEI/10.13039/501100011033 (and the European Union through European Regional Development Fund `A way of making Europe') through grant RTI2018-095076-B-C21, the Institute of Cosmos Sciences University of Barcelona (ICCUB, Unidad de Excelencia `Mar\'{\i}a de Maeztu’) through grant CEX2019-000918-M, the University of Barcelona's official doctoral programme for the development of an R+D+i project through an Ajuts de Personal Investigador en Formaci\'{o} (APIF) grant, the Spanish Virtual Observatory through project AyA2017-84089, the Galician Regional Government, Xunta de Galicia, through grants ED431B-2021/36, ED481A-2019/155, and ED481A-2021/296, the Centro de Investigaci\'{o}n en Tecnolog\'{\i}as de la Informaci\'{o}n y las Comunicaciones (CITIC), funded by the Xunta de Galicia and the European Union (European Regional Development Fund -- Galicia 2014-2020 Programme), through grant ED431G-2019/01, the Red Espa\~{n}ola de Supercomputaci\'{o}n (RES) computer resources at MareNostrum, the Barcelona Supercomputing Centre - Centro Nacional de Supercomputaci\'{o}n (BSC-CNS) through activities AECT-2017-2-0002, AECT-2017-3-0006, AECT-2018-1-0017, AECT-2018-2-0013, AECT-2018-3-0011, AECT-2019-1-0010, AECT-2019-2-0014, AECT-2019-3-0003, AECT-2020-1-0004, and DATA-2020-1-0010, the Departament d'Innovaci\'{o}, Universitats i Empresa de la Generalitat de Catalunya through grant 2014-SGR-1051 for project `Models de Programaci\'{o} i Entorns d'Execuci\'{o} Parallels' (MPEXPAR), and Ramon y Cajal Fellowship RYC2018-025968-I funded by MICIN/AEI/10.13039/501100011033 and the European Science Foundation (`Investing in your future');
\item the Swedish National Space Agency (SNSA/Rymdstyrelsen);
\item the Swiss State Secretariat for Education, Research, and Innovation through the Swiss Activit\'{e}s Nationales Compl\'{e}mentaires and the Swiss National Science Foundation through an Eccellenza Professorial Fellowship (award PCEFP2\_194638 for R.~Anderson);
\item the United Kingdom Particle Physics and Astronomy Research Council (PPARC), the United Kingdom Science and Technology Facilities Council (STFC), and the United Kingdom Space Agency (UKSA) through the following grants to the University of Bristol, the University of Cambridge, the University of Edinburgh, the University of Leicester, the Mullard Space Sciences Laboratory of University College London, and the United Kingdom Rutherford Appleton Laboratory (RAL): PP/D006511/1, PP/D006546/1, PP/D006570/1, ST/I000852/1, ST/J005045/1, ST/K00056X/1, ST/\-K000209/1, ST/K000756/1, ST/L006561/1, ST/N000595/1, ST/N000641/1, ST/N000978/1, ST/\-N001117/1, ST/S000089/1, ST/S000976/1, ST/S000984/1, ST/S001123/1, ST/S001948/1, ST/\-S001980/1, ST/S002103/1, ST/V000969/1, ST/W002469/1, ST/W002493/1, ST/W002671/1, ST/W002809/1, and EP/V520342/1.
\end{itemize}

The GBOT programme  uses observations collected at (i) the European Organisation for Astronomical Research in the Southern Hemisphere (ESO) with the VLT Survey Telescope (VST), under ESO programmes
092.B-0165,
093.B-0236,
094.B-0181,
095.B-0046,
096.B-0162,
097.B-0304,
098.B-0030,
099.B-0034,
0100.B-0131,
0101.B-0156,
0102.B-0174, and
0103.B-0165;
%
%
and (ii) the Liverpool Telescope, which is operated on the island of La Palma by Liverpool John Moores University in the Spanish Observatorio del Roque de los Muchachos of the Instituto de Astrof\'{\i}sica de Canarias with financial support from the United Kingdom Science and Technology Facilities Council, and (iii) telescopes of the Las Cumbres Observatory Global Telescope Network.

\section{Selected RVS spectra\label{sect:selected.spectra}}

\begin{figure*}[!htp]
{\center
\includegraphics[width=0.99\linewidth,clip=,draft=False]{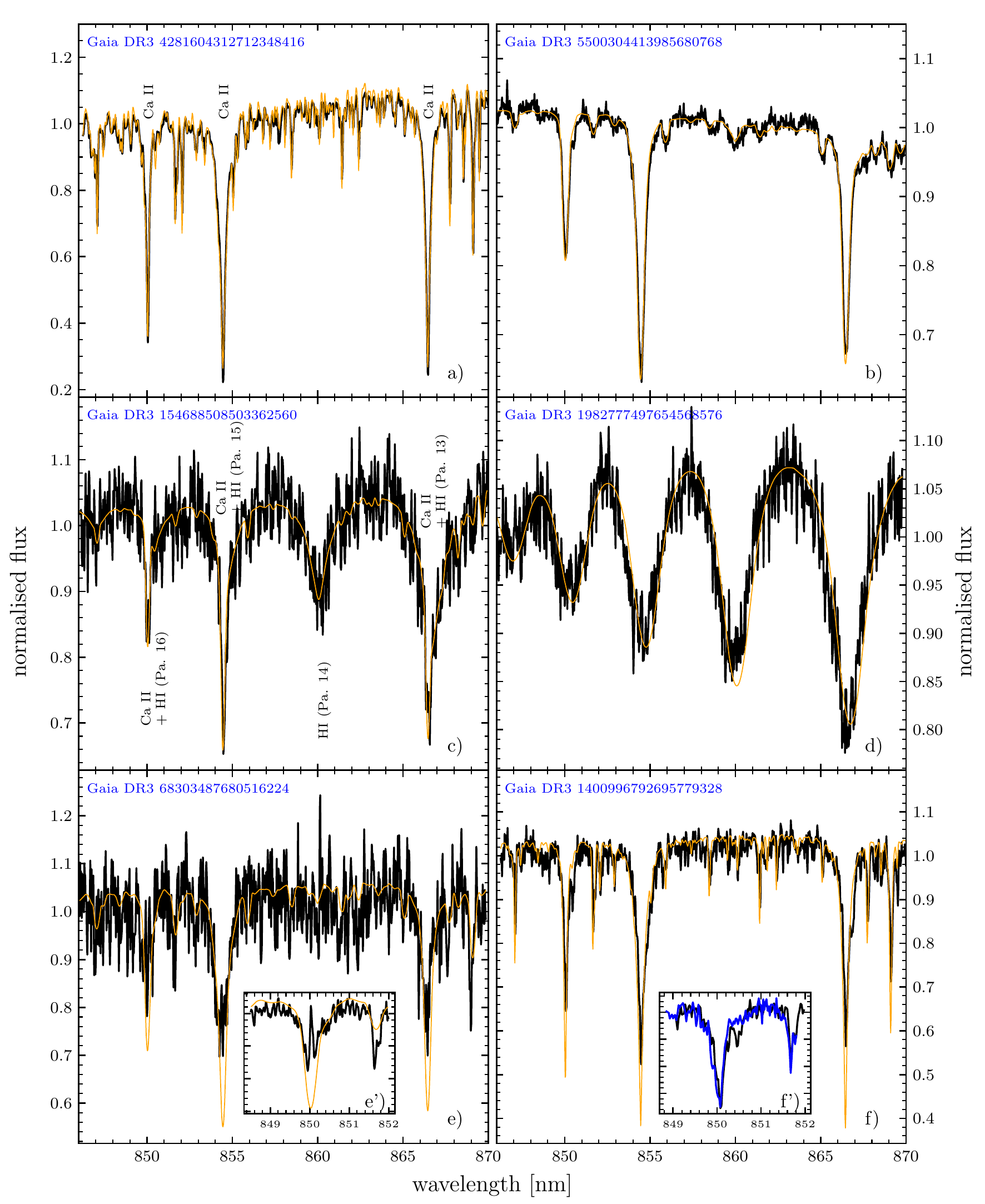}
 \caption{Examples of RVS spectra used to derive the \vbroad\ parameter. Transit spectra (black curve) are compared to the template spectrum used to measure \vbroad\ (orange curve) and broadened to the published estimate. The inset of panel e) zooms in the corresponding {\bf multiple transit combined} spectrum (i.e. black curve in subpanel e') to show the signature of chromospheric activity. The inset of panel f) compares two transit spectra (black and blue) of the same target. The target IDs are given in blue in the panels upper left, while the \grvs\ magnitude, and astrophysical parameters considered to select and broaden the template spectra (orange) are given in Table\,\ref{tab:rvsspectra}. The spectra used to make these plots are not part of the \GDRthree\ release.}\label{fig:rvsspectra}
 }
\end{figure*}

We show in Fig.\,\ref{fig:rvsspectra} a selection of spectra with different values of \grvs, \teff, \logg, \feh, and \vbroad\ (see also Table\,\ref{tab:rvsspectra}). From  cool to hotter targets, panels a) to d) show how the relative strength of the \ion{Ca}{ii} triplet and hydrogen Paschen lines varies with effective temperature. Above $\grvs=10$, the weakest spectral lines, usually more sensitive to \vbroad, are rapidly disappearing in the noise, while in the hottest star (panel d) the main features are the broad lines of the Paschen series. These data are transit spectra, which are not part of the \GDRthree\ release.

The pipeline used to derive the radial velocities is able to flag the most obvious cases of emission-line stars and spectroscopic binaries. However, spectra belonging to targets exhibiting signatures of chromospheric activity (see panel e and its inset) that could not be automatically identified still have published \vbroad\ estimates. The same is true for a fraction of undetected binaries (e.g. those that in most transits are not spectroscopically resolved). One example is presented in panel f) for a target located in the colour magnitude diagram of Fig.\,\ref{fig:HRD} on the binary MS. Line-core emission in the spectra of active stars as well as line-profile asymmetry due to binarity are expected to bias the \vbroad\ determinations.

\begin{table}[!htp]
 \setlength{\tabcolsep}{2.5pt}
    \caption{Description of the template spectra shown in each panel of Fig.\,\ref{fig:rvsspectra}.}
    \label{tab:rvsspectra}
    \centering
    \begin{tabular}{crrrrrr}
\hline
\hline
& & & \multicolumn{3}{c}{{\tt rv\_template\_}} \\
          & \multicolumn{1}{l}{Gaia DR3 ID} & \grvs & {\tt teff} & {\tt logg} & {\tt fe\_h} & {\tt vbroad} \\
& & & [K] & & & [\kps]\\        
\hline
a) & 4281604312712348416 & 5.30 &  3700 & 1.00 & $+$0.25 &   9  \\ 
b) & 5500304413985680768 & 6.78 &  6500 & 4.50 & $-$0.25 &  85  \\
c) &  154688508503362560 & 9.33 &  8000 & 4.50 & $+$0.25 &  54  \\
d) & 1982777497654568576 & 8.66 & 14000 & 4.00 & $+$0.25 & 300  \\
e) &   68303487680516224 & 10.40 &  5500 & 4.50 & $+$0.25 &  73  \\
f) & 1400996792695779328 & 8.88 &  4750 & 5.00 & $+$0.00 &  12  \\
\hline
    \end{tabular}
\end{table}

\end{appendix}
\end{document}